\documentclass[reprint,twocolumn,aps,prl,amsmath,amssymb,floatfix,superscriptaddress,longbibliography]{revtex4-2}
\usepackage{graphicx}
\usepackage{epsfig}
\usepackage{bm}
\usepackage{dcolumn}
\usepackage{color}
\usepackage{physics}
\usepackage{float}
\usepackage[colorlinks,urlcolor=cyan,citecolor=blue,linkcolor=magenta]{hyperref}
\makeatletter
\newcommand*{\rom}[1]{\expandafter\@slowromancap\romannumeral #1@}
\makeatother
\usepackage{tikz}
\usepackage{subfigure}
\usepackage{ulem}
\usepackage{dsfont}
\usepackage{enumitem}
\usepackage{verbatim}
\usepackage{cancel}
\usepackage{ulem}

\newcommand{\ma}{\textcolor[rgb]{0,0.7,0}}

\usepackage{ulem}
\begin{document}

\title{Proposal for Observing Yang-Lee Criticality in Rydberg Atomic Arrays}
\author{Ruizhe Shen}
\email{e0554228@u.nus.edu}
\affiliation{Department of Physics, National University of Singapore, Singapore 117551, Singapore}

\author{Tianqi Chen}
\email{tianqi.chen@ntu.edu.sg}
\affiliation{School of Physical and Mathematical Sciences, Nanyang Technological University, Singapore 639798, Singapore}

\author{Mohammad Mujahid Aliyu}
\email{e0382015@u.nus.edu}
\affiliation{Centre for Quantum Technologies, National University of Singapore, 117543 Singapore, Singapore}

\author{Fang Qin}
\affiliation{Department of Physics, National University of Singapore, Singapore 117551, Singapore}

\author{Yin Zhong}
\affiliation{School of Physical Science and Technology $\&$ Key Laboratory for Magnetism and Magnetic Materials of the MoE, Lanzhou University, Lanzhou 730000, China} %
\affiliation{Lanzhou Center for Theoretical Physics, Key Laboratory of Theoretical Physics of Gansu Province, Lanzhou 730000, China}

\author{Huanqian Loh}
\email{phylohh@nus.edu.sg}
\affiliation{Department of Physics, National University of Singapore, Singapore 117551, Singapore}
\affiliation{Centre for Quantum Technologies, National University of Singapore, 117543 Singapore, Singapore}

\author{Ching Hua Lee}
\email{phylch@nus.edu.sg}
\affiliation{Department of Physics, National University of Singapore, Singapore 117551, Singapore}
\affiliation{Joint School of National University of Singapore and Tianjin University, International Campus of Tianjin University, Binhai New City, Fuzhou 350207, China}
\date{\today}

\begin{abstract}
Yang-Lee edge singularities (YLES) are the edges of the partition function zeros of an interacting spin model in the space of complex control parameters. They play an important role in understanding non-Hermitian phase transitions in many-body physics, as well as characterizing the corresponding nonunitary criticality. 
Even though such partition function zeroes have been measured in dynamical experiments where time acts as the imaginary control field, experimentally demonstrating such YLES criticality with a physical imaginary field has remained elusive due to the difficulty of physically realizing non-Hermitian many-body models. 
We provide a protocol for observing the YLES by detecting kinked dynamical magnetization responses due to broken $\mathcal{P}\mathcal{T}$ symmetry, thus enabling the physical probing of nonunitary phase transitions in nonequilibrium settings. In particular, scaling analyses based on our nonunitary time evolution circuit with matrix product states accurately recover the exponents uniquely associated with the corresponding nonunitary CFT. We provide an explicit proposal for observing YLES criticality in Floquet quenched Rydberg atomic arrays with laser-induced loss, which paves the way towards a universal platform for simulating non-Hermitian many-body dynamical phenomena. 
\end{abstract}
\maketitle

\noindent{\it Introduction.--} 
In 1952, Yang and Lee established a relationship between phase transitions and special points where the partition function vanishes, also known as Yang-Lee zeros~\cite{yang1952statistical,lee1952statistical}. For a spin model in the thermodynamic limit, nonunitary critical points known as Yang-Lee edge singularities (YLES)~\cite{cardy1985conformal,von1991critical,jian2021yang,SuppMat} lie at the ends of a dense line of partition function zeros in the space of complex control parameters such as the magnetic field or inverse temperature~\cite{yang1952statistical,lee1952statistical}. To observe YLES, one can examine a non-Hermitian quantum ferromagnetic many-body Hamiltonian involving complex magnetic fields~\cite{fisher1980yang,matsumoto2020embedding,jian2021yang,SuppMat}. The YLES of such non-Hermitian ferromagnetic models lead to anomalous critical scaling behaviours associated with their governing nonunitary conformal field theories (CFTs)~\cite{yin2017kibble,matsumoto2020embedding,SuppMat}. 

For a long time, Yang-Lee edge singularities, requiring challenging experimental realization of imaginary fields, have been deemed as purely theoretical constructs. Recently, it was realized that the partition function of a classical spin model can also be mathematically simulated by real-time evolution, and Yang-Lee zeros were finally observed through probing spin coherence in a series of landmark experiments involving externally coupled local spins~\cite{peng2015experimental,francis2021many,wei2012lee}.  However, what these experiments achieved was the measurement of partition function zeros through a dynamical process, not the physical observation of the YLES and their associated nonunitary phase transition. It is still difficult to realize such esoteric phenomena in a finite-size quantum ferromagnetic model with physical complex fields. While various single-body non-Hermitian phenomena have already been demonstrated~\cite{ding2021experimental,zhou2021engineering,liang2022dynamic}, experimental demonstrations in {\it interacting} many-body non-Hermitian models have just begun, primarily with cold atoms~\cite{malossi2014full,goldschmidt2016anomalous,li2019observation,ren2022chiral,liang2022observation}.

Indeed, ultracold atomic systems have lately proven to be ideal platforms for simulating many-body physics due to their excellent tunability and high controllability~\cite{bloch2005ultracold,bloch2008many,giorgini2008theory,bloch2012quantum}, with demonstrated successes in topological physics~\cite{miyake2013realizing,jotzu2014experimental}, strongly correlated matter~\cite{greiner2002quantum,mazurenko2017cold,salomon2019direct} and thermalization phenomena~\cite{schreiber2015observation,kyprianidis2021observation}. Rydberg atomic arrays are particularly promising, with Rydberg interactions successfully deployed to simulate many-body Hamiltonians~\cite{barredo2015coherent,labuhn2016tunable,zeiher2017coherent,de2017optical,lin2019exact,browaeys2020many,lin2020quantum,borish2020many,bluvstein2021controlling,geier2021floquet,scholl2022microwave}. Rydberg-dressing techniques offer new possibilities for engineering complex many-body phases due to their great tunability in shaping interactions~\cite{bouchoule2002spin,johnson2010interactions,henkel2010three,pupillo2010strongly,schempp2015correlated,zeiher2016many,jau2016entangling,borish2020transverse,hollerith2022realizing}. To introduce non-Hermiticity in an ultracold atomic lattice, an increasingly established approach~\cite{lee2019coherent,ren2022chiral,li2020topological,bienias2020photon,wintermantel2020unitary,qin2023non,liang2022observation} is laser-induced loss by exciting atoms to ``external'' states~\cite{li2019observation,lapp2019engineering,ren2022chiral}. Encouraged by these recent rapid advances in ultracold atoms, we combine Rydberg-dressing techniques with laser-induced loss to realize a ferromagnetic chain with an imaginary effective field, such as to observe the YLES in a genuinely physical complex parameter space.

In this work, we first introduce a transverse-field Ising model with imaginary fields and discuss its critical  signatures associated with spontaneous $\mathcal{P} \mathcal{T}$ breaking~\cite{giorgi2010spontaneous,galda2018parity,wang2015spontaneous,tuloup2020nonlinearity}.
We next devise a Floquet quench for observing the YLES phase boundary by measuring nonequilibrium kinked responses, before describing our proposed experiment involving a dissipative Rydberg-dressed optical tweezer array~\cite{henkel2010three,pupillo2010strongly,borish2020transverse,schafer2020tools,SuppMat}, where the imaginary field is implemented through laser-induced atom loss~\cite{li2019observation,lapp2019engineering,ren2022chiral}.

\begin{figure}
	\centering
	\includegraphics[width=0.84\linewidth]{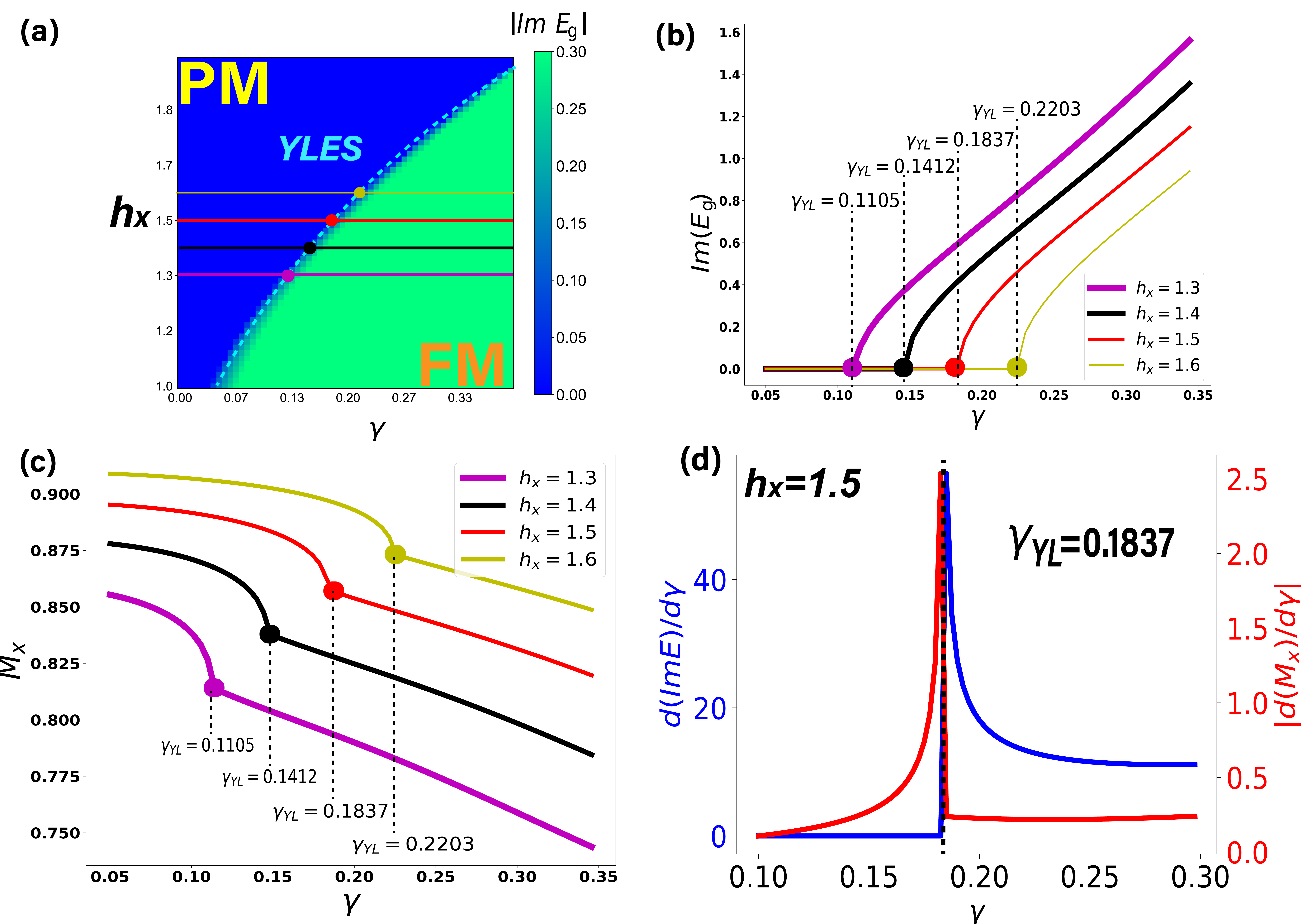}
	\caption{Theoretical characterization of the Yang-Lee edge singularities (YLES) of $\hat H_\text{TFI}$ (Eq.~\eqref{tfi}) and their associated ground state discontinuities: (a) Phase diagram of $\hat{H}_{\rm TFI}$ in the space of real and complex fields $h_x$ and $\gamma$. The YLES (dashed curve) is the phase boundary demarcating the paramagnetic (PM) ${\rm Im}E_{\rm g}=0$ phase and the ferromagnetic (FM) $|{\rm Im}E_{\rm g}|>0$ phase, where $E_g$ is the ground state eigenenergy with the minimal real part.
		(b) Plots of ${\rm Im}E_{\rm g}$ vs. $\gamma$ at the four values of $h_x$ indicated in (a), revealing that $E_g$ is non-analytic at YLES $\gamma_{YL}$.
		(c) Identification of Yang-Lee phase transitions by the ground state magnetization order parameter $M_{x}=\left |\left\langle \sum_{j}\hat{\sigma}^{x}_{j}/L\right\rangle\right|$. Magnetization kinks (dashed lines) occur at the same $\gamma_{YL}$ as in (b). 
		(d) The critical point (at $\gamma_{YL}=0.1837$ for $h_x=1.5$) can be equivalently extracted from divergences in either the derivative of the imaginary ground state energy $\frac{d|{\rm Im}(E_{\rm g}(\gamma))|}{d\gamma}$ (blue), or that of the ground state x magnetization $\frac{dM_{x}}{d\gamma}$ (red). All results are obtained from exact diagonalization (ED) with open boundary conditions, with interaction strength $J=1$ and system size $L=8$.
	}
	\label{fig:1}
\end{figure}
\noindent{\it Model for Yang-Lee edge singularities.--}
For a concrete platform for observing Yang-Lee criticality, we consider the prototypical non-Hermitian ferromagnetic transverse-field Ising chain~\cite{li2014conventional,pires2021probing}, which we will later show the realization in Rydberg atoms:
\begin{equation}\label{tfi}
	\hat{H}_{\rm TFI}=-\sum_{j}(h_{x}\hat{\sigma}_{j}^{x}+J\hat{\sigma}^{z}_{j} \hat{\sigma}^{z}_{j+1})+\sum_{j}i\gamma \hat{\sigma}^{z}_{j},
\end{equation}
where $J$ sets the strength of the interaction, $\gamma$ is the imaginary field strength and $\hat{\sigma}_{j}^{\alpha} (\alpha=x,z)$ are Pauli matrices. The YLES for this model form a curve in the plane of real and imaginary magnetic fields $(h_{x},\gamma)$ in~FIG.~\ref{fig:1} (a)~\cite{SuppMat}, which we denote as $\gamma_{YL}$ at each value of $h_{x}$. When approaching $\gamma_{YL}$, the ground states of our model experience spontaneous $\mathcal{P} \mathcal{T}$ symmetry breaking~\cite{SuppMat}, with real ground-state eigenenergies $E_{g}$ splitting into complex eigenenergies with equal and opposite $\text{Im}E_g$, which demarcate the paramagnetic and ferromagnetic ground states in our model, as shown in FIG.~\ref{fig:1} (a) for $\hat{H}_{\rm TFI}$: the paramagnetic (ferromagnetic) phases are characterized by vanishing (nonvanishing) $\text{Im}E_g$.

Despite YLES initially being defined  for classical systems in the thermodynamic limit, they equivalently exist in finite-size quantum systems due to a quantum-classical mapping~\cite{cardy1985conformal,von1991critical,sachdev1999quantum,sachdev_2011,liu2015universal,silvi2016crossover,yin2017kibble,wei2018probing,SuppMat},
as shown in FIGs.~\ref{fig:1} (a),(b), which were computed through exact diagonalization (ED) with $L=8$ sites. In FIGs.~\ref{fig:1} (c),(d), the associated ground state magnetization $M_{z}=\left |\left\langle \sum_{j}\hat{\sigma}^{z}_{j}/L\right\rangle\right|$ and $M_{x}=\left |\left\langle \sum_{j}\hat{\sigma}^{x}_{j}/L\right\rangle\right|$ also exhibit kinks at these YLES locations. Indeed, comparing the derivatives $\frac{d|({\rm Im}E_{\rm g}(\gamma))|}{d\gamma}$ against that of the x magnetization $|\frac{dM_{x}(\gamma)}{d\gamma}|$ (FIG.~\ref{fig:1} (d)), we observe divergences at the same YLES $\gamma=\gamma_{YL}=0.1837$ (for $J=1,h_x=1.5$).

\noindent{\it Dynamical response from Yang-Lee edge singularities.--}
Since Eq.~\ref{tfi} is non-Hermitian, any physical realization will undergo nonequilibrium evolution, making it difficult to directly probe ground state properties such as Yang-Lee criticality via a static ensemble measurement. A nonequilibrium detection of ground state transitions requires different behaviors across phases, in particular with one dominating the dynamics~\cite{dora2019kibble,xiao2021non,haldar2021signatures,SuppMat}.

\begin{figure}
	\centering
	\includegraphics[width=.85\linewidth]{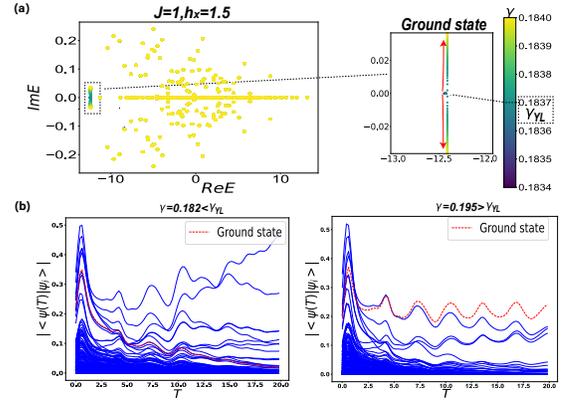}
	\caption{(a) Flow of the complex spectrum of $\hat H_\text{TFI}$ (Eq.~\eqref{tfi}) as $\gamma$ is tuned across $\gamma_{YL}\approx 0.1837$. The ground state eigenenergies (boxed) undergo spontaneous $\mathcal{P} \mathcal{T}$ symmetry breaking and rapidly acquire imaginary parts (green to yellow, red arrow) as $\gamma$ increases slightly above $\gamma_{YL}$, much more so than other eigenenergies.  
		(b) and (c): Evolution of the overlap $|\bra{\psi(T)}\ket{\psi_{i}}|$ between the dynamically evolved state $\ket{\psi(T)}\!=\!\hspace{-1mm}e^{-iT\hat{H}_{\rm TFI}}\ket{\psi(0)}/||e^{-iT\hat{H}_{\rm TFI}}\ket{\psi(0)}||$ and all eigenstates $\ket{\psi_{i}}$ (blue) of $\hat{H}_{\rm TFI}$, with initial state being $\ket{\psi(0)}=\ket{\downarrow\downarrow......}$. For $\gamma$ below the YLES $\gamma_{YL}$ (b), the ground state overlap (red dashed) decreases rapidly due to mixing. But for $\gamma>\gamma_{YL}$, the ground state component becomes dominant beyond time $TJ\sim \mathcal{O}(10)$ due to the large $\text{Im}E_g$ of the ground state, leading to kinked magnetization responses detectable in our proposed Rydberg atom system of FIG.~\ref{fig:ising} below. All results are obtained from ED with $J=1$, $h_x=1.5$ and $L=8$.
	}
	\label{fig:spec}
\end{figure}

To probe the YLES dynamically, we turn to the spectral flow across critical points. The underappreciated but crucial observation is that due to $\mathcal{PT}$-symmetry breaking, imaginary eigenenergies appear and that leads to markedly different nonunitary dynamics across the transition~\cite{SuppMat}. From FIG.~\ref{fig:spec} (a), the ground state energy $E_g$ (with smallest $\text{Re}E_g$) is seen to rapidly acquire larger $\pm \text{Im}E_g$ immediately after $\gamma$ is tuned to be greater than $\gamma_{YL}$ (inset). By contrast, other nonground state eigenenergies in FIG.~\ref{fig:spec} (a) are largely stationary. This drastic real-to-complex ground state eigenenergy transition does not just imply that ground state observables i.e. $M_x$ exhibit a kink at the YLES - more importantly, the rapidly increasing $\text{Im}E_g$ at $\gamma>\gamma_{YL}$ suggests that upon time evolution by $\hat H_\text{TFI}$, {\it any} initial state with significant ground state overlap will converge towards the ground state and dominate.

As such, our proposal to experimentally detect nonunitary critical YLES involves measuring the dynamical $x$-magnetization order parameter $M_{x}(T)\!=\!\bra{\psi(t)}\sum_{j}\hat{\sigma}^{x}_{j}/L\ket{\psi(t)}$, which we henceforth expect to exhibit similar kinks as the ground state $M_x$ (FIG.~\ref{fig:1} (c)). The edge of such nonunitary phase transitions can be identified by plotting the kink locations of $M_x(T)$ in parameter space. With our Rydberg array implementation in mind, we propose the following protocol:
\begin{enumerate}
	\item Prepare an ordered initial state $\ket{\psi(0)}=\ket{\downarrow\downarrow......}$. \footnote{In our proposed Rydberg array, we initialize the  $\ket{\psi(0)}=\ket{\leftarrow\leftarrow\leftarrow......}_y$ state for a better implementation of interactions, but the two initial states are both eligible \cite{SuppMat}.}
	\item Apply quench dynamics on this ordered initial state, resulting in normalized result: $\ket{\psi(t)}\!=\!\hspace{-1mm}e^{-it\hat{H}_{\rm TFI}}\ket{\psi(0)}/||e^{-it\hat{H}_{\rm TFI}}\ket{\psi(0)}||$ with $||\ket{\psi}||=\sqrt{|\langle \psi | \psi \rangle|}$.
	\item Measure the x magnetization order parameter $M_x(T)\!=\!|\bra{\psi(T)}\sum_{j}\hat{\sigma}^{x}_{j}\ket{\psi(T)}|/L$ after a sufficiently long stipulated time $T$ for different $\gamma$, keeping $J$ and $h_{x}$ fixed.
\end{enumerate}
As shown in FIGs.~\ref{fig:spec} (b),(c), the initial ferromagnetic state $|\psi(0)\rangle$ already overlaps with the ground state of $\hat H_\text{TFI}$ more than most other eigenstates. Because of the PT-symmetry breaking, after evolving for $TJ\sim \mathcal{O}(10^1)$, it is dominated by the ground state for $\gamma>\gamma_{YL}$, but not when $\gamma<\gamma_{YL}$. As such, we expect to observe a kink in $M_x(T)$ across the critical YLES $\gamma=\gamma_{YL}$, even though there is considerable ground state overlap at one side of the transition. Note that even though the ground state $M_x(T)$ is extracted through a dynamical quench, what we are measuring is not mathematically a dynamical phase transition \cite{peng2015experimental,francis2021many,wei2012lee,brandner2017experimental}. Instead, our proposal for measuring nonunitary criticality can be demonstrated by our designed experimental optical arrays discussed below.

\begin{figure}
	\centering
	\includegraphics[width=1\linewidth]{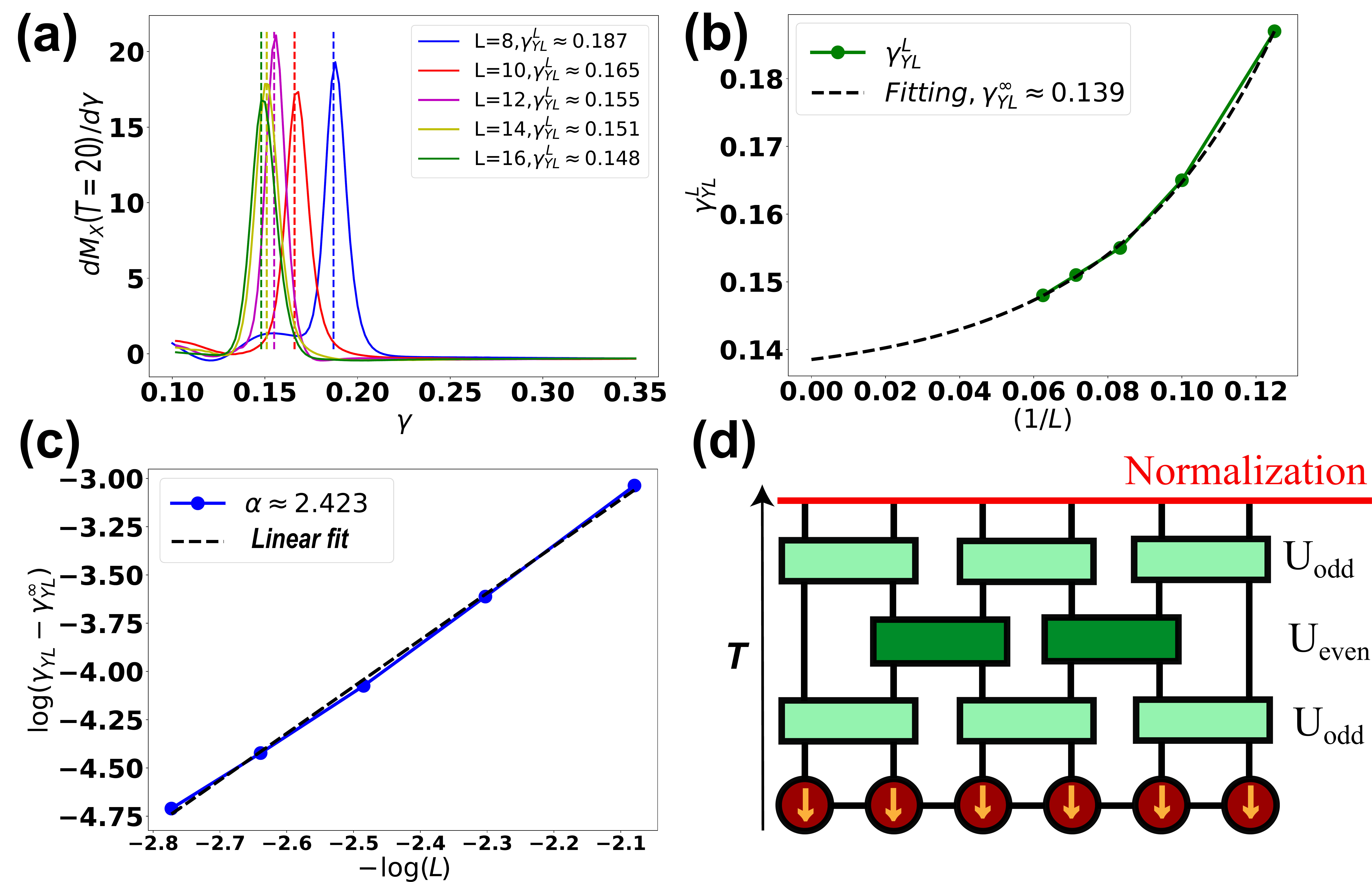}
	\caption{Anomalous critical scaling of dynamically determined Yang-Lee edge singularities (YLES) of $\hat H_\text{TFI}$, with $J=1$ and $h_x=1.5$. (a) Size-dependent YLES $\gamma_{YL}^L$ as kinks in the derivative $\frac{dM_{x}(T)}{d\gamma}$. The dynamical magnetization $M_x(T)=|\bra{\psi(T)}\sum_{j}\hat{\sigma}^{x}_{j}\ket{\psi(T)}|/L$ is measured at time $T=20/J$ for different system sizes $L=8$ to $16$ using our tMPS algorithm from (d). 
		(b) Extrapolation of $\gamma_{YL}^{L}$ with $1/L$ gives the thermodynamic limit YLES value $\gamma_{YL}^{\infty}\approx0.139$ through polynomial fitting.	
		(c) Plotting our data for $\log(\gamma_{YL}^{L}-\gamma_{YL}^{\infty})$ against $-\log L$ yields the critical exponent $\alpha\approx2.423$ as the gradient (Eq.~\eqref{scal}), which agrees closely with the CFT result $\alpha=12/5$~\cite{fisher1978yang,uzelac1979yang}. 	
		(d) The nonunitary circuit for implementing the time evolution of our model~\cite{SuppMat}, where each step is decomposed into odd-bond $U_{\text{odd}}$ (light green) and even-bond $U_{\text{even}}$ (dark green) parts, with normalization (red solid line) performed at the end.
	}
	\label{fig:mx}
\end{figure}

\noindent{\it Nonunitary criticality of Yang-Lee edge singularities.--} We then demonstrate how to determine the YLES critical exponents using finite-size scaling with our protocol.  After fixing the quenching duration $T$, the critical values $\gamma=\gamma^L_{YL}$ at different $L$ can be extracted from the peak divergences of the plots of the derivative of $M_{x}(T)$ with respect to $\gamma$. This is illustrated in  FIG.~\ref{fig:mx} (a) with data from our tMPS simulation (described shortly after), computed with $T=20/J$ with $J=1$~\cite{SuppMat}. Here, for different system sizes $L=8,10,...,16$, the critical values of $\gamma=\gamma^L_{YL}$ are marked by dashed vertical lines, where the derivatives peak. In an experiment, $\gamma^L_{YL}$ can be extracted by adjusting $\gamma$ and measuring $M_x(T)$ in separate spin chains of various lengths $L$. 

The characteristic critical exponents for the YLES can be extracted via the following universal critical scaling law of its nonunitary $c=-22/5$ conformal field theory (CFT)~\cite{von1993multi,yin2017kibble}:
\begin{equation}\label{scal}
	\begin{aligned}
		\gamma^{L}_{YL}-\gamma^{\infty}_{YL}\propto L^{-\alpha}= L^{-\left(\beta_{1} \delta_{1} / \nu_{1}\right)},	\end{aligned}
\end{equation}
where $\gamma^{L}_{YL}$ is the location of the YLES obtained from our protocol at finite size $L$. $\gamma^{\infty}_{YL}$ represents the YLES as $L\rightarrow\infty$, which can be obtained from our finite-sized data by extrapolating $\gamma^L_{YL}$ with respect to $1/L$, as performed by polynomial fitting in FIG.~\ref{fig:mx} (b).
Upon obtaining $\gamma_{YL}^\infty$, one can further plot $\log(\gamma^{L}_{YL}-\gamma^{\infty}_{YL})$ against $\log(1/L)$, such that the critical exponent $\alpha$ in Eq.~\ref{scal} can be extracted from the gradient of the fitted line in FIG.~\ref{fig:mx} (c). Such nonunitary CFT behavior, which has been elusive in experiments, can be readily measured in our proposed cold atom setup. Our tMPS result $\alpha_{\rm MPS} \approx 2.423$ is in excellent agreement with the theoretical value from the nonunitary CFT with central charge $c=-22/5$ for the YLES: $\alpha_{\rm CFT}=\beta_{1} \delta_{1} / \nu_{1}=2.40$ with $\beta_{1}=1,\delta_{1}=-6,\nu_{1}=-5/2$~\cite{fisher1978yang,uzelac1979yang}.

Before discussing the detailed experimental process, we briefly describe our designed nonunitary numerical algorithm for the results in FIG.~\ref{fig:mx}, which is built upon the state-of-art matrix product states (tMPS) tool for handling generic one-dimensional quantum many-body systems with nearest-neighbor couplings~\cite{Suzuki1990,White1992,vidal2004efficient,White2004}. As sketched in FIG.~\ref{fig:mx} (d) \footnote{In this work, we only consider open boundary conditions, which is less costly for MPS calculations.}, we build a nonunitary circuit for $e^{-i\delta t \hat{H}_{\rm TFI}}$ ~\cite{gingrich2004non,koczor2019quantum,xie2022probabilistic,SuppMat}, which is implemented through a second-order Suzuki-Trotter decomposition built by nonunitary even and odd bonds~\cite{SuppMat}. A key feature is that normalization at the end of each time step suppresses numerical divergences, facilitating the precise computation of long-time dynamics ($T=30$)~\cite{SuppMat}. Therefore, our method offers direct, efficient implementations, outperforming ancilla-based approaches with significant information wastage and poor scalability~\cite{gingrich2004non,daskin2017ancilla}. 

\begin{figure}
	\centering
	\includegraphics[width=0.98\linewidth]{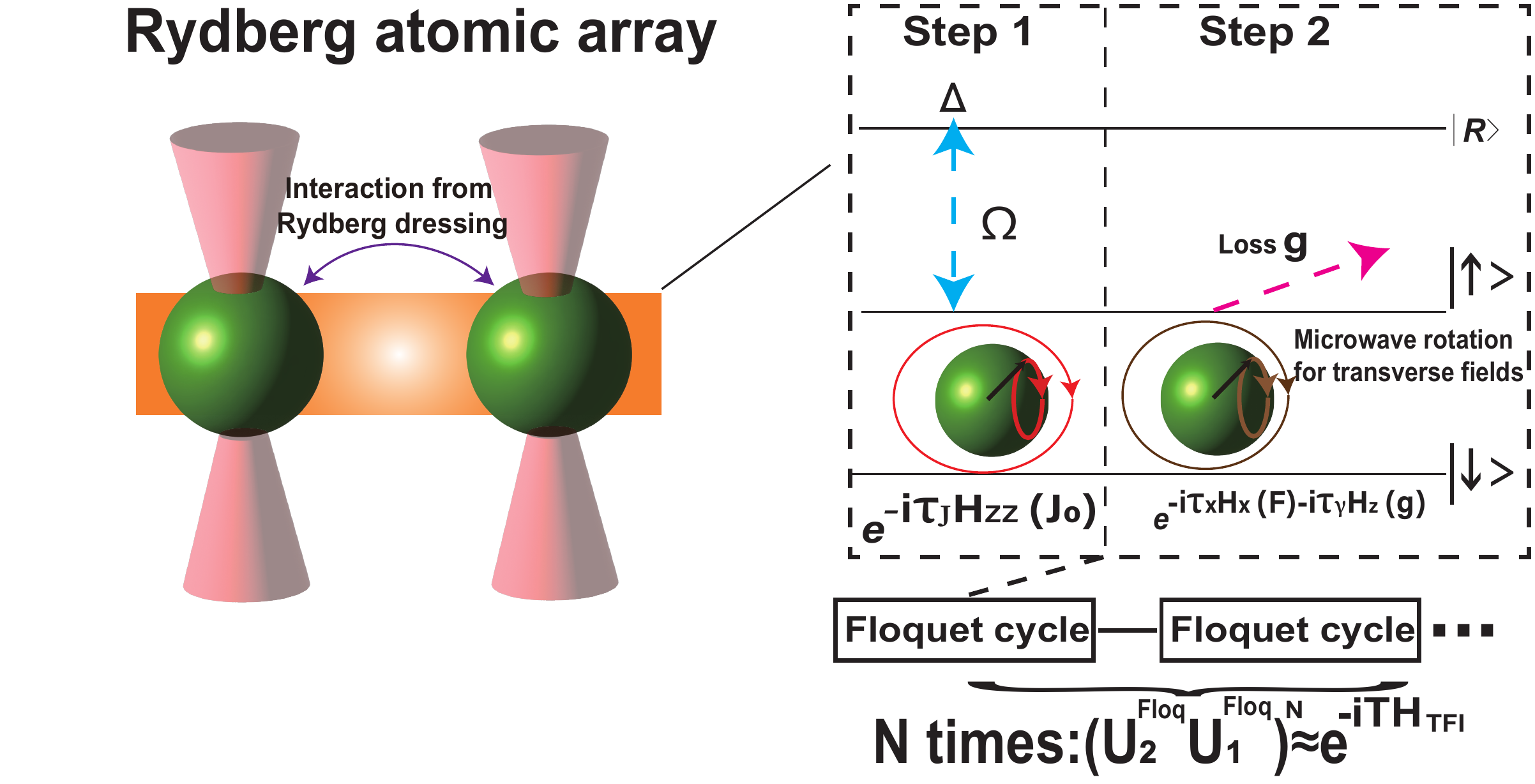}
	\caption{Proposed Rydberg atomic array for realizing the Floquet drives $U_1^\text{Floq}$ and $U_2^\text{Floq}$ in Eq.~\eqref{cycle}. Optical tweezers (red, Left) trap Cesium atoms (green), with their two hyperfine ground states defining pseudospin-$1/2$ states $\ket{\downarrow}\hspace{-1mm}=\hspace{-1mm}\left|6 S_{1 / 2}, F\hspace{-1mm}=\hspace{-1mm}3, m_{F}\hspace{-1mm}=\hspace{-1mm}+3\right\rangle$ and $\ket{\uparrow}\hspace{-1mm}=\hspace{-1mm}\left|6 S_{1 / 2}, F\hspace{-1mm}=\hspace{-1mm}4, m_{F}\hspace{-1mm}=\hspace{-1mm}+4\right\rangle$ (right). In $U_1^\text{Floq}$ (Step 1), $\Omega$ driving (blue dashed arrow) couples $\ket{\uparrow}$ and the Rydberg state $\ket{60 P_{3 / 2}, m_J = +3 /2}$, resulting in a Rydberg-dressed nearest-neighbor $\ket{\uparrow\uparrow}$ interaction $\hat H_\text{int}$ (purple arrow) that can be converted to a ferromagnetic interaction via spin echo (red loop). In $U_2^\text{Floq}$ (Step 2), a microwave field (brown loop) generates the transverse field $F\hat{\sigma}^{x}$, while a laser couples $\ket{\uparrow} \rightarrow\ket{6P_{3/2}, F = 5, m_F = +5}$, resulting in the imaginary field $ig\hat{\sigma}^{z}$~\cite{SuppMat}. The quenching durations $\tau_{J}$, $\tau_{x}$ and $\tau_{\gamma}$ are chosen such that $(\tau_{x}F,\tau_{\gamma}g,\tau_{J}J_0)=\frac{T}{N}(h_{x},\gamma,J)$.}
	\label{fig:ising}
\end{figure}

\noindent{\it Experimental proposal with Rydberg atoms.--}
We then show how to implement the dynamics $e^{-iT\hat{H}_{\rm TFI}}$ to observe the YLES through our driving protocol in Rydberg atomic arrays. Given that in a Rydberg system \cite{borish2020transverse,ren2022chiral}, the implementation of transverse fields $\hat H_X(F)=-\sum_jF\hat\sigma_j^x$, atomic loss $\hat H_Z(g)=-i\sum_j g\hat \sigma_j^z$  and, ferromagnetic interactions $\hat H_{ZZ}=-\sum_{j} J_{0}\hat{\sigma}_{i}^{z}\hat{\sigma}_{i+1}^{z}$ are all feasible, we can Trotterize $e^{-i{T}\hat{H}_{\rm TFI}}$ into $N$ periods by a two-step Floquet driving protocol (FIG.~\ref{fig:ising})
\begin{equation}\label{cycle}
	\begin{aligned}
		e^{-i{T}\hat{H}_{\rm TFI}}&\approx 
\left[e^{+i\tau_{x}\hat{H}_{X}(\!F\!)+i\tau_{\gamma}\hat{H}_{Z}(\!g\!)}e^{+i\tau_{J}\hat{H}_{ZZ}}\right]^N\\	
		&=[U_2^\text{Floq}U_1^\text{Floq}]^N,	\end{aligned}
\end{equation}
where $U_1^\text{Floq}=e^{i\tau_{J}\hat{H}_{ZZ}}$ followed by $U_2^\text{Floq}=e^{i\tau_{x}\hat{H}_{X}(\!F\!)+i\tau_{\gamma}\hat{H}_{Z}(\!g\!)}$, and the physical magnitudes $F$, $g$ and $J_{0}$ are related to their corresponding quenching durations $\tau_{J}$, $\tau_{x}$ and $\tau_{\gamma}$ via $(\tau_{x}F,\tau_{\gamma}g,\tau_{J}J_0)=\frac{T}{N}(h_{x},\gamma,J)$. The Floquet cycles $N=\frac{TJ}{J_{0}\tau_{J}}$ needs to be determined under the condition of coherent dynamics~\footnote{In this limit of Trotterization, the first order correction can sometimes already give rise to interesting new phenomena~\cite{vcadevz2017dynamical,li2018realistic,rudner2019floquet,nathan2021quasiperiodic,castro2022floquet,qin2022light}, particularly in many-body settings~\cite{kuwahara2016floquet,lee2018floquet,ye2021floquet}. Details of Floquet cycles are shown in \cite{SuppMat} and Refs. \cite{steck2007quantum,macri2016loschmidt,jau2016entangling,mcdonnell2022demonstration,graham2022,hines2023spin}.}. 

We next detail the experimental Floquet quenches $U_1^\text{Floq}$ and $U_2^\text{Floq}$ in our Rydberg setup, built by the pseudospin-1/2 of two hyperfine Caesium ground states: $\ket{\downarrow}\hspace{-1mm}=\hspace{-1mm}\left|6 S_{1 / 2}, F\hspace{-1mm}=\hspace{-1mm}3, m_{F}\hspace{-1mm}=\hspace{-1mm}+3\right\rangle$ and $\ket{\uparrow}\hspace{-1mm}=\hspace{-1mm}\left|6 S_{1 / 2}, F\hspace{-1mm}=\hspace{-1mm}4, m_{F}\hspace{-1mm}=\hspace{-1mm}+4\right\rangle$ (FIG.~\ref{fig:ising}) \cite{SuppMat}.
To engineer interactions in the first step $U_1^\text{Floq}$, we couple the state $\ket{\uparrow}$  with the Rydberg state $\ket{60 P_{3 / 2}, m_J = +3 /2}$ under large detuning $\Delta$ vs. Rabi frequency $\Omega$ (blue arrow in FIG.~\ref{fig:ising})\cite{SuppMat}. Consequently, the Rydberg-dressing process leads to an energy shift of $J_{0}\approx \frac{\Omega^{4}}{8\Delta^{3}}$ in the nearest-neighbor subspace $\ket{\uparrow\uparrow}_{i,i+1}$, resulting in effective interactions~\cite{johnson2010interactions,balewski2014rydberg,SuppMat} 
\begin{align}\label{ising}
	\hat{H}_{\rm int}(J_{0})=-\sum_{i} J_{0}\hat{P}_{i}\hat{P}_{i+1},
\end{align}
where $\hat{P}_{i}=\frac{\hat{\sigma}_{i}^{z}+\hat{I}_{i}}{2}$ with $\hat{\sigma}^{z}_i=\ket{\uparrow}_i\bra{\uparrow}_i-\ket{\downarrow}_i\bra{\downarrow}_i$ and $\hat{I}_i=\ket{\uparrow}_i\bra{\uparrow}_i+\ket{\downarrow}_i\bra{\downarrow}_i$. While this is still not the ferromagnetic interaction required in $\hat H_\text{TFI}$ of Eq.\eqref{cycle}, inspired by the operator identity $e^{-i\frac{\pi}{2}\hat{\sigma}^{x}}e^{i\tau_{J} {J_{0}}\hat{\sigma}^{z}}e^{-i\frac{\pi}{2}\hat{\sigma}^{x}}e^{i\tau_{J} {J_{0}}\hat{\sigma}^{z}}=-\hat{I}$, one can convert it into a clean ferromagnetic interaction $\hat{H}_{ZZ}(J_{0})=-\sum_{j} J_{0}\hat{\sigma}_{i}^{z}\hat{\sigma}_{i+1}^{z}$, by applying two transverse $\hat{\sigma}^{x}$ field kicks:
\begin{equation}\label{zz}
	\begin{aligned}
		(e^{-i\frac{\pi}{2}\sum_{j}\hat{\sigma}_{j}^{x}}e^{i\tau_{J}\hat{H}_{\rm int}(2J_{0})})^{2}\approx U_1^\text{Floq}, 
	\end{aligned}
\end{equation}
as elaborated in the Supplemental Material~\cite{SuppMat}. Such transverse-field kicks can be generated by microwave fields, shown as the red circle in FIG.~\ref{fig:ising}~\cite{borish2020transverse}. 

To realize the next evolution step $U_2^\text{Floq}$, the Rydberg dressing for $U_1^\text{Floq}$ is turned off immediately, and another microwave field (brown circle in FIG.~\ref{fig:ising}) is turned on to generate the transverse field $\hat{H}_{X}(F)=-\sum_{j}F\hat{\sigma}_{j}^{x}$ \cite{borish2020transverse}. At the same time, a strong laser is shone on $\ket{\uparrow}$ such as to excite it to another state $\ket{6P_{3/2},F=5}$, leading to effective laser-induced loss $\hat{H}_{Z}(g)=\sum_{j}ig\hat{\sigma}_{j}^{z}$ with imaginary field/decay rate $g$~\cite{saffman2010quantum,lee2019coherent,bienias2020photon,wintermantel2020unitary,lourencco2022non,SuppMat}.

After repeatedly alternating between Floquet steps $U_1^\text{Floq}$ and $U_2^\text{Floq}$ (FIG.~\ref{fig:ising}) over $N=\frac{TJ}{J_0\tau_J}$ iterations, the dynamically evolved magnetization $M_x(T)$ can be obtained by measuring the normalized populations in the $\ket{\uparrow}$ and $\ket{\downarrow}$ levels~\cite{zhou2021engineering,SuppMat}. The YLES can be observed as kinks in $M_x(T)$ as $\tau_{\gamma}$ or $g$ are tuned (Fig.~\ref{fig:mx}). From that, the associated anomalous scaling behavior and exponents (Eq.~\ref{scal}) can be simply be extracted by controlling the number of trapped atoms $L$~\cite{browaeys2020many,scholl2021quantum}.

{\noindent{\it Discussions.--}}
The ground state properties of non-Hermitian quantum systems are often deemed experimentally inaccessible due to overwhelming decoherence or the lack of thermal equilibrium. Yet, for the Yang-Lee phase transition in our model, we found that the spontaneously broken $\mathcal{PT}$-symmetry can give rise to pronounced kinks in the dynamical magnetization $M_x(T)$ {\it without} the need for reaching thermal equilibrium. As such, we provide a realistic Floquet evolution protocol for observing the YLES criticality in a Rydberg chain, distinct from the observation of partition function zeroes in previous experiments~\cite{peng2015experimental,francis2021many,wei2012lee}. Our proposal paves the way for future experimental observation of not just the YLES, but also other nonunitary phase transitions~\cite{zhou2018dynamical,wang2019observation,hamazaki2021exceptional,zhang2022universal,lin2022topological,lourencco2022non}.

The rapid development of universal quantum computation also opens up the possibility of implementing our YLES measurement protocol in quantum computers via ancilla-based methods~\cite{lin2021real,koh2022stabilizing,koh2022simulation,chen2022high,koh2023observation}. Moreover, we achieved precise long-time dynamics through our MPS approach for nonunitary circuits. This approach, which is related to mid-circuit measurements~\cite{pino2021demonstration,egan2021fault,bharti2022noisy,hua2022exploiting}, offers a route to improve the current ancilla-based methods for dynamically simulating various non-Hermitian many-body phenomena~\cite{le2013steady,sieberer2013dynamical,joshi2013quantum,landa2020multistability,jian2020criticality,pistorius2020quantum,deuar2021fully,naji2022dissipative} and unconventional non-Hermitian topology~\cite{okuma2020topological,lee2021many,zhai2022nonequilibrium,li2021quantized,li2022non,shen2022non,fu2022anatomy,lee2022exceptional,kawabata2022many,alsallom2022fate,qin2022universal,yoshida2023fate,poddubny2023interaction,guo2023non}, on quantum circuits.

\begin{acknowledgements}
{\noindent{\it Acknowledgements.--}} T.~C. thanks Bo Yang for discussions. The exact diagonalization is  computed with QuSpin Python library~\cite{weinberg2017quspin,weinberg2019quspin}, and MPS results are calculated with ITensor~\cite{itensor}. T.~C. acknowledges support from the National Research Foundation, Singapore under the NRF fellowship award (NRF-NRFF12-2020-005). F.~Q. is supported by the QEP2.0 grant from the Singapore National Research Foundation (Grant No.~NRF2021-QEP2-02-P09) and the MOE Tier-II grant (Proposal ID T2EP50222-0008). All data and code of this work are available from the corresponding authors upon reasonable request.
\end{acknowledgements}

\bibliography{references}
\newpage
\appendix
\onecolumngrid
\setcounter{equation}{0}
\setcounter{figure}{0}
\setcounter{table}{0}
\setcounter{section}{0}
\renewcommand{\theequation}{S\arabic{equation}}
\renewcommand{\thefigure}{S\arabic{figure}}
\renewcommand{\thesection}{S\arabic{section}}
\section*{Supplementary Materials for ``Proposal for Observing Yang-Lee Criticality in Rydberg Atomic Arrays''}
In this supplement, we provide the following:

I. Elaboration on engineering a non-Hermitian transverse-field Ising model in a Rydberg atomic array.

II. Phase transitions associated with Yang-Lee edge singularities.

III. Details of our tMPS numerical simulation approach and results.

\section{I. Rydberg atom array platforms}
In the main text, we have suggested observing the Yang-Lee edge singularities (YLES) in a non-Hermitian transverse-field Ising model, which is proposed in a Rydberg atom array platform. In Rydberg atom arrays, the Rydberg-dressing technique can generate ferromagnetic interactions among ground-state atoms~\cite{henkel2010three,pupillo2010strongly,zeiher2017coherent,borish2020transverse}, and atom loss leads to imaginary fields~\cite{li2019observation,lapp2019engineering,ren2022chiral}. In this section, we further provide more details for our experimental proposal.
\subsection{Atomic energy levels}
Realizing our proposal of observing the YLES  requires implementing  state-selective, laser-induced atom loss, and interactions among atoms. Therefore, at least one of the qubit states need to form a closed cycling transition with another atomic level to avoid having the atoms optically pumped into a dark state after a few scattering events \ma{\cite{steck2007quantum}}. For this reason, we propose to use $\ket{\uparrow} = |6S_{1/2}, F=4, m_F=+4\rangle$, and $\ket{\downarrow} = |6S_{1/2}, F=3, m_F=+3\rangle$ in Cesium, where the $\ket{\uparrow} \rightarrow|6P_{3/2}, F=5, m_F=+5\rangle$ forms a closed cycling transition for the laser-induced atom loss discussed later. The Rydberg state for engineering interactions is chosen to be $\ket{R} = \ket{60P_{3/2},m_J = +3/2}$, which has a blackbody radiation-limited lifetime of $\tau_{ryd} = 148~\mu$s (natural linewidth $\Gamma = 2\pi \times 1.1~$kHz) and interaction parameter of $C_{6} = 2\pi \times 360$~GHz . By avoiding the F\"{o}rster resonance at the $n \approx 43$ Rydberg state, we can avoid additional decoherence mechanisms \cite{hines2023spin}.

\subsection{Atoms with Rydberg dressing}
To engineer interactions among atoms in our setup by Rydberg-dressing techniques~\cite{henkel2010three,pupillo2010strongly,zeiher2017coherent,borish2020transverse}, we couple the ground state $\ket{\uparrow}$ with the Rydberg state $\ket{R}$ as shown FIG 1 of the main text. Note that the ground state here refers to the atomic energy level, which is different from the eigenstate of our transverse-field Ising model in Eq.\eqref{supptfi} discussed later. To illustrate how such a coupling process leads to effective interactions, we consider a simple two-atom case, where we couple the Rydberg state $\ket{R}$ with the ground state $\ket{\uparrow}$ by means of a Rabi frequency $\Omega$~\cite{johnson2010interactions}. This setup can be described by a Hamiltonian in the basis of $(|\uparrow \uparrow\rangle, 1 / \sqrt{2}(|\uparrow R\rangle+|R \uparrow\rangle),|RR\rangle)$:
\begin{equation}
	\hat{H}_{\rm dressing}=\left(\begin{array}{ccc}
		0 & \frac{\Omega}{\sqrt{2}} & 0 \\
		\frac{\Omega}{\sqrt{2}} & \Delta & \frac{\Omega}{\sqrt{2}} \\
		0 & \frac{\Omega}{\sqrt{2}} & 2 \Delta+U_{d d}
	\end{array}\right),
\end{equation}
where $\Delta$ is laser detuning at the ground state $\ket{\uparrow}$, and $U_{dd}$ is for dipole-dipole interaction between two nearest Rydberg states~\cite{johnson2010interactions}. We can approximate the ground-state energy of $\hat{H}_{\rm dressing}$ as 
\begin{equation}
	E_{G}\approx\frac{1}{2}\left(\Delta-\sqrt{2 \Omega^{2}+\Delta^{2}}\right)
\end{equation}
in the strong interaction limit as $U_{dd}\gg\Delta$~\cite{johnson2010interactions}. Compared with the ground-state energy without Rydberg dressing $E_{G}|_{U_{dd}=0}=\Delta-\sqrt{\Omega^{2}+\Delta^{2}}$, we obtain an energy shift on states $\ket{\uparrow\uparrow}$, which in the large detuning limit $\Delta\gg\Omega$ is
\begin{equation}\label{int}
	\delta E_{G}\approx-\frac{\Omega^{4}}{8\Delta^{3}},
\end{equation}
We propose to use the parameters provided in Table~\ref{tab:dressingparams} to implement the Rydberg dressing.

\begin{table}[htb] 
	\caption{Proposed dressing parameters.} \label{tab:dressingparams}
	\begin{center}
		\begin{tabular}{|l|c|}
			\hline
			Dressing laser linewidth, $\gamma_{l}$ & $2\pi \times$ 7~kHz \\
			Dressing laser Rabi frequency, $\Omega$ & $2\pi \times$ 6.8~MHz \\ 
			Dressing laser detuning, $\Delta$ & $2\pi \times$ 22~MHz \\
			Separation between atoms, $r$ & 3.4~$\mu$m \\
			Interaction range, $r_c = (\frac{C_6}{2\Delta})^{1/6}$ & 4.5~$\mu$m \\
			Peak interaction, $U_0$ = $\frac{\Omega^4}{8\Delta^3}$ & 2$\pi\times$25.9~kHz \\
			Interaction strength at $r$, $U(r)$ = $\frac{U_0}{1 + (r/r_c)^6}$ & $2\pi \times 22.0$~kHz \\
			Coupling strength, $J_{0}(r) = U(r)/4$ & $2\pi \times$5.5~kHz \\
			%\hline 
			\hline
		\end{tabular}
	\end{center}
\end{table}
\subsection{Kicks from transverse fields}
As discussed in the main text and under `Floquet quenches', the Rydberg-dressing process produces the effective interaction  as follows
\begin{equation}\label{supising}
	\hat{H}_{\rm int}(J_{0})=-J_0\sum_i \ket{\uparrow\uparrow}_{i,i+1}\bra{\uparrow\uparrow}_{i,i+1}=-\sum_{i} \frac{J_{0}}{4}\left(\hat{\sigma}_{i}^{z}+\hat{I}_{i}\right)\left(\hat{\sigma}_{i+1}^{z}+\hat{I}_{j}\right),
\end{equation}
with $J_{0}\hspace{-1.5mm}=\hspace{-1.5mm}\Omega^{4} /\left|8 \Delta^{3}\right|$. Here, we  neglect the dressing laser-induced light shift, and the next-nearest-neighbour interactions are not significant~\cite{johnson2010interactions}. 

To realize clean ferromagnetic interactions, $\hat{H}_{ZZ}(J_{0})=-\sum_{i} J_{0}\hat{\sigma}_{i}^{z}\hat{\sigma}_{i+1}^{z}$, as required for the transverse field Ising model, spin echo techniques are generally employed to remove the unwanted terms after polarizing the atoms in the equatorial plane of the Bloch sphere~\cite{zeiher2017coherent}. Using the identity in Eq.(\ref{supkick}), we show that the Hamiltonian in Eq.(\ref{supising}) can equally be transformed to the desired Ising interactions required for realizing the transverse field Ising model.
\begin{equation}\label{supkick}
	e^{-i\frac{\pi}{2}\hat{\sigma}^{x}}e^{+i\delta \tau {J_{0}}\hat{\sigma}^{z}}e^{-i\frac{\pi}{2}\hat{\sigma}^{x}}e^{+i\delta \tau {J_{0}}\hat{\sigma}^{z}}=\left(\begin{array}{cc}
		-1 & 0 \\
		0 & -1
	\end{array}\right).
\end{equation} 

Physically, the exponential terms involving $\hat \sigma^x$ can be interpreted as two kicks from transverse $\hat{\sigma}^{x}$ fields. The fact that the right-hand-side of Eq.~\ref{supkick} is proportional to the identity means that the kicks can remove the $e^{+i\delta \tau {J_{0}}\hat{\sigma}^{z}}$ field.

When generalized to the whole array, such kicks can also eliminate the intermediate $\hat{\sigma}^{z}$-field contributions in Eq.~\ref{supising}, such as to convert $\hat H_\text{int}$ to a clean ferromagnetic interaction. To describe this explicitly, we first expand the fast-evolving operation, 
\begin{equation}\label{zz}
	%\begin{aligned}
	U_1^\text{Floq} = (e^{-i\frac{\pi}{2}\sum_{j}\hat{\sigma}_{j}^{x}}e^{+i\tau_{J}\hat{H}_{\rm int}(2J_{0})})^{2}, 
	%\end{aligned}
\end{equation}

dropping errors of subleading order in $\delta \tau$~\cite{borish2020transverse}:
\begin{equation}\label{suppcycle}
	\begin{aligned}
		&(e^{-i\frac{\pi}{2}\sum_{j}\hat{\sigma}_{j}^{x}}e^{+i\delta\tau\sum_{u} \frac{J_{0}}{2}\left(\hat{\sigma}_{u}^{z}+\hat{I}_{u}\right)\left(\hat{\sigma}_{u+1}^{z}+\hat{I}_{u+1}\right)})^{2}\\
		&=(e^{-i\frac{\pi}{2}\sum_{j}\hat{\sigma}_{j}^{x}}e^{+i\delta \tau(\sum_{u} \frac{J_{0}}{2}\sigma_{u}^{z}\hat{\sigma}_{u+1}^{z})})(e^{+i\sum_{j}\delta \tau {J_{0}}\hat{\sigma}_{j}^{z}}e^{-i\frac{\pi}{2}\sum_{j}\hat{\sigma}_{j}^{x}}e^{+i\sum_{j}\delta \tau {J_{0}}\hat{\sigma}_{j}^{z}})(e^{+i\delta \tau(\sum_{u} \frac{J_{0}}{2}\hat{\sigma}_{u}^{z}\hat{\sigma}_{u+1}^{z})})+\mathcal{O}(\delta \tau^2).
	\end{aligned}
\end{equation}
Then, by mathematically inserting the $\hat{\sigma}^{z}$-field kicks as $e^{+i\frac{\pi}{2}\sum_{j}\hat{\sigma}_{j}^{x}}e^{-i\frac{\pi}{2}\sum_{j}\hat{\sigma}_{j}^{x}}$, we can write the above process as follows
\begin{equation}\label{suppzz1}
	\begin{aligned}
		&(e^{-i\frac{\pi}{2}\sum_{j}\hat{\sigma}_{j}^{x}}e^{+i\delta \tau(\sum_{u} \frac{J_{0}}{2}\sigma_{u}^{z}\hat{\sigma}_{u+1}^{z})}e^{+i\frac{\pi}{2}\sum_{j}\hat{\sigma}_{j}^{x}})(e^{-i\frac{\pi}{2}\sum_{j}\hat{\sigma}_{j}^{x}}e^{+i\sum_{j}\delta \tau {J_{0}}\hat{\sigma}_{j}^{z}}e^{-i\frac{\pi}{2}\sum_{j}\hat{\sigma}_{j}^{x}}e^{+i\sum_{j}\delta \tau {J_{0}}\hat{\sigma}_{j}^{z}})(e^{+i\delta \tau(\sum_{u} \frac{J_{0}}{2}\hat{\sigma}_{u}^{z}\hat{\sigma}_{u+1}^{z})}).
	\end{aligned}
\end{equation}		
Since $(e^{-i\frac{\pi}{2}\sum_{j}\hat{\sigma}_{j}^{x}}e^{+i\sum_{j}\delta \tau {J_{0}}\hat{\sigma}_{j}^{z}}e^{-i\frac{\pi}{2}\sum_{j}\hat{\sigma}_{j}^{x}}e^{+i\sum_{j}\delta \tau {J_{0}}\hat{\sigma}_{j}^{z}})$ can be removed due to the operator identity in Eq.\eqref{supkick}, the following sequence of driving terms can recover the clean ferromagnetic interaction $\hat{H}_{ZZ}(J_{0})=-\sum_{i} J_{0}\hat{\sigma}_{i}^{z}\hat{\sigma}_{i+1}^{z}$:
\begin{equation}\label{suppzz2}
	\begin{aligned}		
		&(e^{-i\frac{\pi}{2}\sum_{j}\hat{\sigma}_{j}^{x}}e^{+i\delta \tau(\sum_{u} \frac{J_{0}}{2}\sigma_{u}^{z}\hat{\sigma}_{u+1}^{z})}e^{+i\frac{\pi}{2}\sum_{j}\hat{\sigma}_{j}^{x}})(e^{+i\delta \tau(\sum_{u} \frac{J_{0}}{2}\hat{\sigma}_{u}^{z}\hat{\sigma}_{u+1}^{z})})\\
		&\approx e^{+i\delta\tau\sum_{u} {J_{0}}\hat{\sigma}_{u}^{z}\hat{\sigma}_{u+1}^{z}},
	\end{aligned}
\end{equation}
with $e^{-i\frac{\pi}{2}\sum_{j}\hat{\sigma}_{j}^{x}}e^{+i\delta \tau(\sum_{u} \frac{J_{0}}{2}\sigma_{u}^{z}\hat{\sigma}_{u+1}^{z})}e^{+i\frac{\pi}{2}\sum_{j}\hat{\sigma}_{j}^{x}}=e^{+i\delta \tau(\sum_{u} \frac{J_{0}}{2}\sigma_{u}^{z}\hat{\sigma}_{u+1}^{z})}.$ Thus, we can write Eq.~\ref{zz} as follows
\begin{equation}\label{zz2}
	%\begin{aligned}
	U_1^\text{Floq} = e^{+i\delta \tau(\sum_{u} J_{0}\hat{\sigma}_{u}^{z}\hat{\sigma}_{u+1}^{z})}.
	%\end{aligned}
\end{equation}

\subsection{Atom loss as imaginary fields}
In our proposed setup, we consider laser-induced atom loss obtained by resonantly driving the $\ket{\uparrow} \rightarrow \ket{6 P_{3/2}, F = 5, m_{F} = 5}$ transition. This gives rise to effective imaginary fields that can be presented as:
\begin{equation}\label{loss}
	\begin{aligned}
		i2g\ket{\uparrow}\bra{\uparrow}&=ig(\ket{\uparrow}\bra{\uparrow}-\ket{\downarrow}\bra{\downarrow})+ig(\ket{\uparrow}\bra{\uparrow}+\ket{\downarrow}\bra{\downarrow})\\
		&=ig\hat{\sigma}^{z}+ig\hat{I},
	\end{aligned}
\end{equation}
where $g$ is the tunable decay rate controlled by laser. Here, we can drop the global loss term $ig\hat{I}$.

\subsection{Experiment sequence}
In this section, we discuss details of our proposed experimental implementation.
Our devised protocol comprises three parts: state initialization, Floquet quenches, and measurement of the magnetization.  Here we elaborate on the experiment sequence in each part.

\subsubsection{State initialization}
First,  all atoms are trapped in optical tweezers separated by 3.4~$\mu$m. The atoms are initialized in the state $\ket{\uparrow\uparrow\uparrow \ldots \uparrow}$ by using a $\sigma^+$-polarized optical pumping beam on the $D_1$ transition \cite{mcdonnell2022demonstration, jau2016entangling}. We note that the magnetic field, which sets the quantization axis, must be perpendicular to the long axis of atom array, so as to maximize the interaction range between the Rydberg-dressed atoms. Then, a microwave pulse to can be used to rotate the spins about the $\hat{S}_{x}$ axis to initialize them in a state fully magnetized along $\hat{S}_{y}$: $R_x(\frac{\pi}{2})\ket{\uparrow\uparrow\uparrow...\uparrow}$ = $\ket{\leftarrow\leftarrow\leftarrow...\leftarrow}_{y}$, where $\ket{\leftarrow}_{y}$ = $(|\uparrow\rangle-i|\downarrow\rangle)/\sqrt{2}$ and $R_x(\frac{\pi}{2})=e^{-\frac{i\pi}{2}\sum_{j}\hat{\sigma}_{j}^{x}}$ \cite{macri2016loschmidt,borish2020transverse,zeiher2016many,zeiher2017coherent}. We point out that this initial state, $\ket{\leftarrow\leftarrow\leftarrow...\leftarrow}_{y}$, also results in similar critical points that are required to characterize the Yang-Lee phase transition as the comparison in FIG.~\ref{fig:mxmz} shows.

\subsubsection{Floquet quenches}\label{ss:Floquet}
\begin{figure}[!htbp]
	\centering
	\includegraphics[keepaspectratio,scale=0.82]{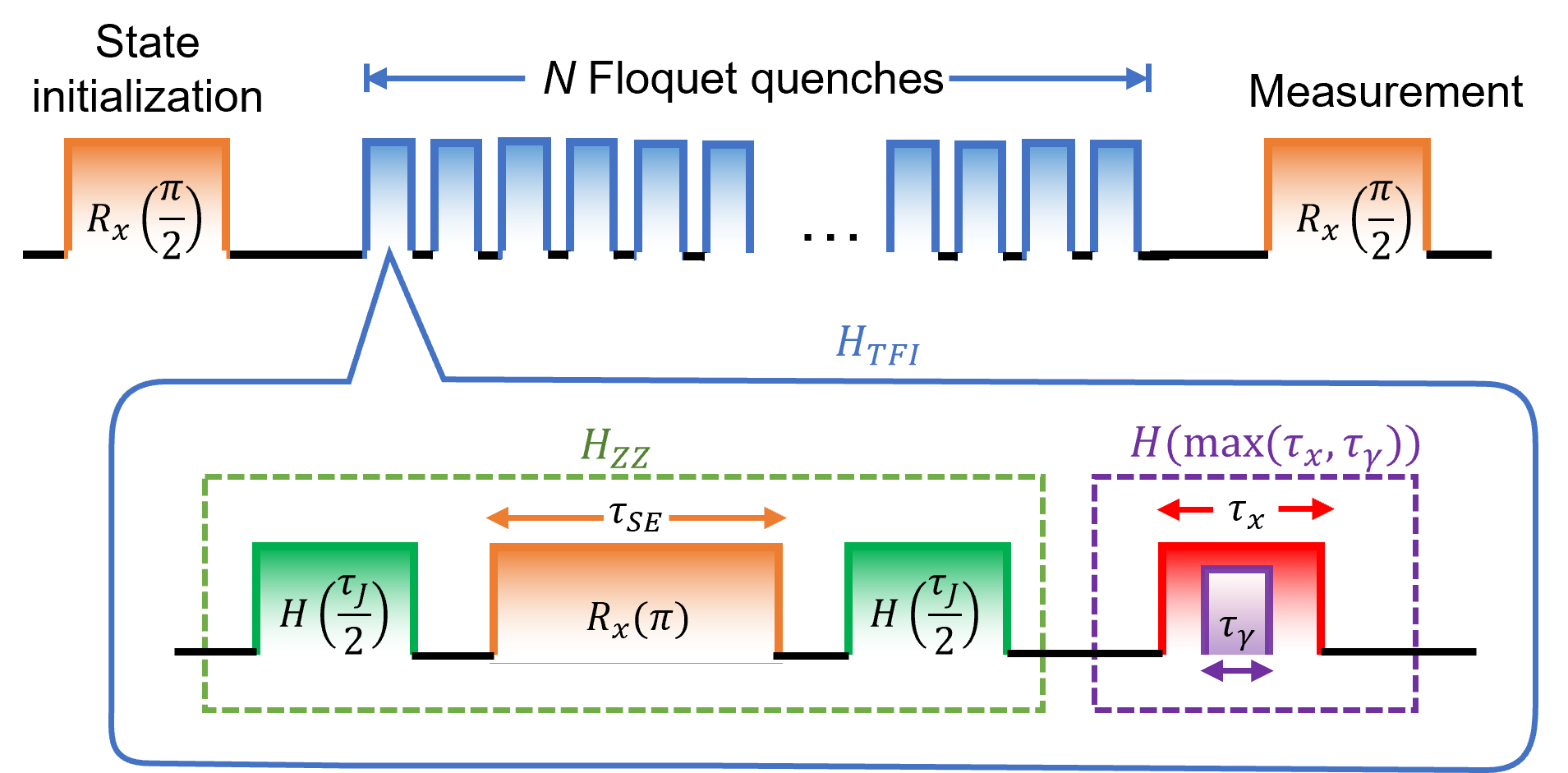}
	\caption{Proposed experiment sequence. The first $R_x(\frac{\pi}{2})$ pulse initializes the state $\ket{\leftarrow\leftarrow\leftarrow...\leftarrow}_{y}$, which then undergoes $N$ Floquet quenches before its magnetization components are detected with another $R_x(\frac{\pi}{2})$ pulse. Each Floquet quench comprises two steps, the first with $H_{ZZ}$ achieved by inserting a spin-echo pulse of duration $\tau_{SE}$ in between two Rydberg-dressing pulses of duration $\tau_{J}/2$, and the second with a transverse field and laser-induced loss to achieve imaginary fields. The final $R_{x}(\frac{\pi}{2})$ is optionally applied if one desires to measure $M_{y}$ or $M_{x}$ but would be omitted for a measurement of $M_{z}$.} 
	\label{fig:floquetquench}
\end{figure}

Here we elaborate on the Floquet sequence for realizing the dynamics of the following model
\begin{equation}\label{supptfi}
	\hat{H}_{\rm TFI}=-\sum_{j}(h_{x}\hat{\sigma}_{j}^{x}+{J}\hat{\sigma}^{z}_{j} \hat{\sigma}^{z}_{j+1})+\sum_{j}i{\gamma} \hat{\sigma}^{z}_{j}.
\end{equation}
As discussed in the main text, each Floquet quench comprises two steps $U_1^\text{Floq}$ and $U_2^\text{Floq}$, the first of which performs the Rydberg dressing, whereas the second step simultaneously implements a transverse field $\hat H_X(F)=-\sum_jF\hat\sigma_j^x$ and an imaginary field $\hat H_Z(g)=-i\sum_j g\hat \sigma_j^z$. The whole process is shown in FIG.~\ref{fig:floquetquench}.

The first step requires using spin echoes to transform the Hamiltonian that is native to Rydberg dressing (Eq.~\ref{eq:nativeH}) to a pure Ising Hamiltonian \cite{borish2020many}:
\begin{eqnarray}\label{ryd}
	H &=& U_{AC}\sum_i(\sigma_i^z + 1/2) + \sum_{i<j}U(r_{ij})(\sigma_i^z + 1/2)(\sigma_j^z + 1/2) \nonumber \\
	&=& \overbrace{\frac{U_{AC}}{2} + \sum_{i<j}J_{0}(r_{ij})}^{\text{offset}} + \overbrace{\sum_{i<j}(\frac{U_{AC}}{2}\sigma_i^z + J_{0}(r_{ij})(\sigma_i^z + \sigma_j^z))}^{\text{$\sigma^z$-terms}} + \overbrace{\sum_{i<j}J_{0}(r_{ij})\sigma_i^z\sigma_j^z}^{\text{Ising}} \, ,
	\label{eq:nativeH}
\end{eqnarray}
where U$_{AC}$ is the dressing laser-induced light shift, and $J_{0}(r_{ij}) = U(r_{ij})/4$. The \textit{offset} terms only add an overall shift to all the states and can be ignored since we only measure relative energies. The \textit{$\sigma_z$} terms, on the other hand, rotate the $x$/$y$ magnetization components around the $\hat{z}$ axis and can cause significant dephasing. Including the spin echo in Eq.~\eqref{zz} allows us to completely remove these terms. Finally, the \textit{Ising} term is the only one that remains after removing the offset and $\sigma_z$ terms, as desired. In our work, the Rydberg dressing parameters are chosen such that only the nearest neighbour interactions are significant, see the section on 'Dissipation'.

The $H_{ZZ}$ block depicted in FIG.~\ref{fig:floquetquench} shows the proposed Floquet pulse sequence with the dressing step split into two pulses, each of duration $\tau_J/2$, in between which the spin echo pulse $R_{x}(\pi)$ of duration $\tau_{SE}$ is inserted. This block corresponds to the $U_1^{Floq}$ evolution in the main text. We note that as long as $\tau_J, \tau_{SE} \ll \tau_{Ryd}$, the spin echo in Eq.~\eqref{zz} will be able to remove any incoherent excitations to the Rydberg state that manifests as a $\sigma_z$-dependent term.

Immediately after the $H_{ZZ}$ block, we simultaneously apply both the transverse field $\hat H_X(F)=-\sum_jF\hat\sigma_j^x$ (for duration $\tau_{x}$) and laser-induced atom loss $\hat H_Z(g)=-i\sum_j g\hat \sigma_j^z$ (for duration $\tau_{\gamma}$) for a time $t$ = max($\tau_{x}$, $\tau_{\gamma}$). This corresponds to $U_2^{Floq}$ in the main text, and thereby completes one Floquet cycle.

The Floquet quench is repeated a total of $N$ times before the magnetization components are read out. In general, we require a large number of Floquet cycles, e.g.\ $N = 50$. For sufficient evolution time $T$, we require $T \times J \geq 10$. Setting $T \times J = 10$ and $N = \frac{TJ}{J_{0} \tau_{J}} = 50$ yields $\tau_J = 1/(5J_{0}) = 5.8~\mu$s. We then note that for a reasonable microwave Rabi frequency of $\Omega_{MW} = 2\pi \times 70$~kHz, $\tau_{SE} = \pi/\Omega_{MW} = 7.1~\mu$s. Choosing $\tau_{x} = 3~\mu$s and $\tau_{\gamma} = 1.5~\mu$s, we get the total evolution time to be $T = N(\tau_J + \tau_{SE} + \text{max}(\tau_{hx}, \tau_\gamma)) \approx 800~\mu$s.

Here we show that these can be realized reasonably:
\begin{itemize}
	\item The constrain of evolution time $T \times J = 10$ yields $J = 2\pi \times 2.0$~kHz. The relevant dressing Rabi frequency and detuning are already provided in Table~\ref{tab:dressingparams}.
	\item According to FIG.~1 (a) in the main text,   $h_{x} = 1.5J = 2\pi \times 3.0$~kHz. Note that $h_x \tau_{x} \ll 1$ as required by the Floquet quench to be effective. One can set $F = (T/\tau_{x}) \times (h_x/N) = 2\pi\times$ 16~kHz. The required microwave Rabi frequency is $2F$, which can be easily achieved \cite{graham2022}.
	\item According to FIG.~1 (b) in the main text, $\gamma$ needs to be increased up to $0.35J = 2\pi \times 0.70$~kHz. Note that $\gamma \tau_{\gamma} \ll 1$ is fulfilled.    The corresponding value of the  atom loss is $g = (T/\tau_{\gamma}) \times (\gamma/N) = 2\pi\times 7.4$~kHz. This can be achieved by using a laser beam of waist 5~mm, optical power $2.4~\mu$W and tuned to resonance with the $\ket{\uparrow} \rightarrow \ket{6P_{3/2}, F = 5, m_{F} = +5}$ cycling transition, such that $g = 2\Gamma_{diss}$ where $\Gamma_{diss}$ corresponds to the laser-induced scattering rate.
\end{itemize}

\subsubsection{Measurement}
At the readout stage, $M_z$ = $\frac{1}{N}\sum_{i}\langle S_i^{z} \rangle$ can be determined by immediately applying spin-sensitive imaging after the Floquet quenches (i.e.\ without the final $R_x(\frac{\pi}{2})$ pulse pictured in Fig.~\ref{fig:floquetquench}). Spin-sensitive detection can be performed by clearing out the atoms in $\ket{\uparrow}$ using a pushout beam resonant with the $D_2$ $\ket{\uparrow} \rightarrow |{F' = 5, m_F' = 5}\rangle$  cycling transition in the absence of a repumper, and then detecting the remaining atoms in $\ket{\downarrow}$. We note that the pushout beam can be the same beam used for performing laser-induced atom loss in the Floquet quench step, except here the pushout beam will operate at a higher intensity.

To measure $M_y$ = $\frac{1}{N}\sum_{i}\langle S_i^{y} \rangle$, we apply a phase-coherent $\frac{\pi}{2}$ pulse after the Floquet quenches to map $\ket{\leftarrow\leftarrow\leftarrow...\leftarrow}_{y}$ to $\ket{\downarrow\downarrow\downarrow...\downarrow}$. For $M_x$, the $\frac{\pi}{2}$ pulse would have to be shifted by a $\frac{\pi}{2}$ phase shift relative to the state-initialization pulse.

\subsection{Dissipation}
We note that the dissipative effects observed in Ref.~\cite{borish2020transverse} and the reduced coherence and short lifetimes observed in Ref.~\cite{zeiher2016many} were attributed to collective effects due to the presence of multiple atoms within the interaction range ($r_c$), which can be incoherently excited to the Rydberg state as a result of the finite linewidth of the dressing laser. These effects were shown to be nearly absent in a 1D system with nearest neighbour interactions \cite{zeiher2017coherent}, where the finite range of the interaction potential ensures the absence of collective decay channels. In particular, the parameters proposed in Table~\ref{tab:dressingparams} are chosen such that nearest-neighbour (NN) interactions dominate ($\frac{U_{NNN}}{U_{NN}} = 8\%$ for r = 3.4~$\mu$m, r$_c$ = 4.5~$\mu$m).

Nevertheless, even with nearest neighbor interactions being dominant, there is a fundamental limit on observing coherent Rydberg-dressed interactions, which is set by the ratio of the coherent coupling term $J_0$ to the effective decay rate of the dressed state $\gamma_{\textit{eff}}$. The latter is given by half the dressed state linewidth due to the fact that the atoms are initialized in $\ket{\leftarrow \leftarrow \leftarrow \ldots \leftarrow}_y$ as opposed to $\ket{\uparrow \uparrow \uparrow \ldots \uparrow}$:
\begin{equation*}
	\gamma_{\textit{eff}} = \frac{\Gamma}{4} \left(\frac{s_0}{1+s_{0} + (2\Delta/\Gamma)^2}\right) = 2\pi \times 0.013~\text{kHz} \, ,
\end{equation*}
where $s_0 = 2\Omega^2/\Gamma^2$ with $\Omega=2\pi \times$ 6.8~MHz, and $\Gamma= 2\pi \times 1.1~$kHz denotes the Rydberg state linewidth. Rewriting $J_0/\gamma_{\textit{eff}}$ as $\Omega^2/(2\Delta\Gamma)$ and observing that $\Omega/(2\Delta) \ll 1$ for the Rydberg dressing to be in the perturbative limit, we require the following condition to be met for coherent Rydberg-dressed interactions \cite{borish2020transverse}:
\begin{equation}
	\frac{J_0}{\gamma_{\textit{eff}}} \ll \frac{\Omega}{\Gamma} \, .
	\label{eq:fundamentallimit}
\end{equation}
According to our proposed parameters, we find that $J_0/\gamma_{\textit{eff}} \sim 430$ and $\Omega/\Gamma \sim$ 6200, allowing us to fulfill Eq.~(\ref{eq:fundamentallimit}).

We recall that any incoherent excitation to the Rydberg state can be quickly removed by the spin echo, as long as $\frac{\tau_{SE}}{2\tau_{ryd}}\ll 1$ and $\frac{\tau_J}{2\tau_{ryd}} \ll 1$. We check that the proposed parameters so far yield $\frac{\tau_{SE}}{2\tau_{ryd}} = 0.024$ and $\frac{\tau_J}{2\tau_{ryd}} = 0.020$ with  $\tau_{ryd} = 148~\mu$s, thereby fulfilling these conditions. Finally, if $\gamma_l\tau_J/2 \ll 1$ cannot be fulfilled, where $\gamma_l=2\pi\times7$~kHz is for the Dressing laser linewidth, we have to factor in decay arising from the finite laser linewidth \cite{zeiher2016many, borish2020transverse}. Our proposed parameters allow us to achieve $\gamma_l \tau_J/2 = 0.13$, so we neglect this source of dissipation for now.

\section{II. Phase transitions in the non-Hermitian transverse-field Ising model}
The Yang-Lee theorem indicates that phase transitions are associated with zeros in partition functions called the Yang-Lee zeros~\cite{yang1952statistical,lee1952statistical}. For a classical spin model in the thermodynamic limit, the Yang-Lee edge singularities correspond to edges of the Yang-Lee zeros~\cite{fisher1980yang}. For our proposed quantum model used in the main text,
the Yang-Lee theorem is also applicable in understanding its quantum phase transitions~\cite{von1991critical,jian2021yang} via a quantum-classical mapping, as described below~\cite{sachdev1999quantum}. 

\section{Quantum-classical mapping}

To understand the existence of the Yang-Lee edge singularities in our finite size model, we consider its partition function $\mathcal{Z}=\operatorname{Tr} \exp (-\beta H_{\rm TFI})$ with inverse temperature $\beta$. Following the quantum-classical mapping in Ref.~\cite{sachdev_2011}, we decompose this partition function expanded into $N_{\beta}\rightarrow \infty$ slices ordered by imaginary time:
\begin{equation}\label{pf}
	\begin{aligned}
		\operatorname{Tr} \exp (-\beta \hat{H}_{\rm TFI})&=\sum _{\sigma_{i}=\uparrow,\downarrow}\bra{\left\lbrace \sigma_{i}\right\rbrace }\exp(\beta\sum_{j=1}^{L}(h_{x}\hat{\sigma}_{j}^{x}+{J}\hat{\sigma}^{z}_{j} \hat{\sigma}^{z}_{j+1})-\sum_{j=1}^{L}i{\gamma} \sigma^{z}_{j})\ket{\left\lbrace {\sigma}_{i}\right\rbrace }\\
		&=\lim_{N_{\beta}\rightarrow\infty}\sum _{\sigma_{i,j}=\pm 1}C\exp(\frac{\beta}{N_{\beta}}(\sum_{i=1}^{L}\sum_{j=1}^{N_{\beta}}({J}\sigma_{i+1,j}\sigma_{i,j}+h^{\prime}\sigma_{i,j}\sigma_{i,j+1})-\sum_{i=1}^{L}\sum_{j=1}^{N_{\beta}}i{\gamma}\sigma_{i,j})),
	\end{aligned}
\end{equation}
with $ \exp (-2 h^{\prime})=\tanh (\frac{\beta}{N_{\beta}}h_{x} )$. Here, we express the partition function of our quantum Ising chain in the basis of product state as $\ket{\left\lbrace \sigma_{i}\right\rbrace }=\prod_{i=1}^{L}\ket{\sigma_{i}}$. For the classical model, $\sigma_{i,j}$ denotes a classical spin, where the coordinate $j$ runs along slices in imaginary time. Note that the anisotropic couplings in the classical spin model do not impact phase transitons, and details of the above mapping are discussed in Ref.~\cite{sachdev_2011}.  In this case, the quantum phase transition of our model will occur when the zeros of its partition function appear under the condition $\beta\rightarrow \infty$. Thus, we have mapped the partition function of a finite-size (1+1) D quantum spin chain to that of a (2+1) D classical spin model in the thermodynamic limit, with the additional dimension arising from the discretization of the imaginary time dimension. With the above quantum-classical mapping, critical points of the Hamiltonian in Eq.~\eqref{supptfi} can be explained as the YLES in the dual classical spin model~\cite{von1991critical,cardy1985conformal,wei2018probing,jian2021yang}.
\subsection{Phase transitions of the finite quantum model}
To determine the phase transitions of our model in Eq.~\eqref{supptfi} (but not that of the dynamical quench in the main text), we introduce the following order parameters:
\begin{equation}\label{order}
	\begin{aligned}
		M_x=|\left\langle \hat{\sigma}^{x}\right\rangle|=\left |\left\langle \frac{\sum_{j}\hat{\sigma}^{x}_{j}}{L}\right\rangle\right |,
	\end{aligned}
	~
	\begin{aligned}
		M_{z}=|\left\langle \hat{\sigma}^{z}\right\rangle|=\left|\left\langle \frac{\sum_{j}\hat{\sigma}^{z}_{j}}{L}\right\rangle\right|,
	\end{aligned}
\end{equation}
which are computed in the ground state with the minimum real eigenvalue.
\begin{figure}[h!]\label{X}
	\centering
	\includegraphics[width=0.65\linewidth]{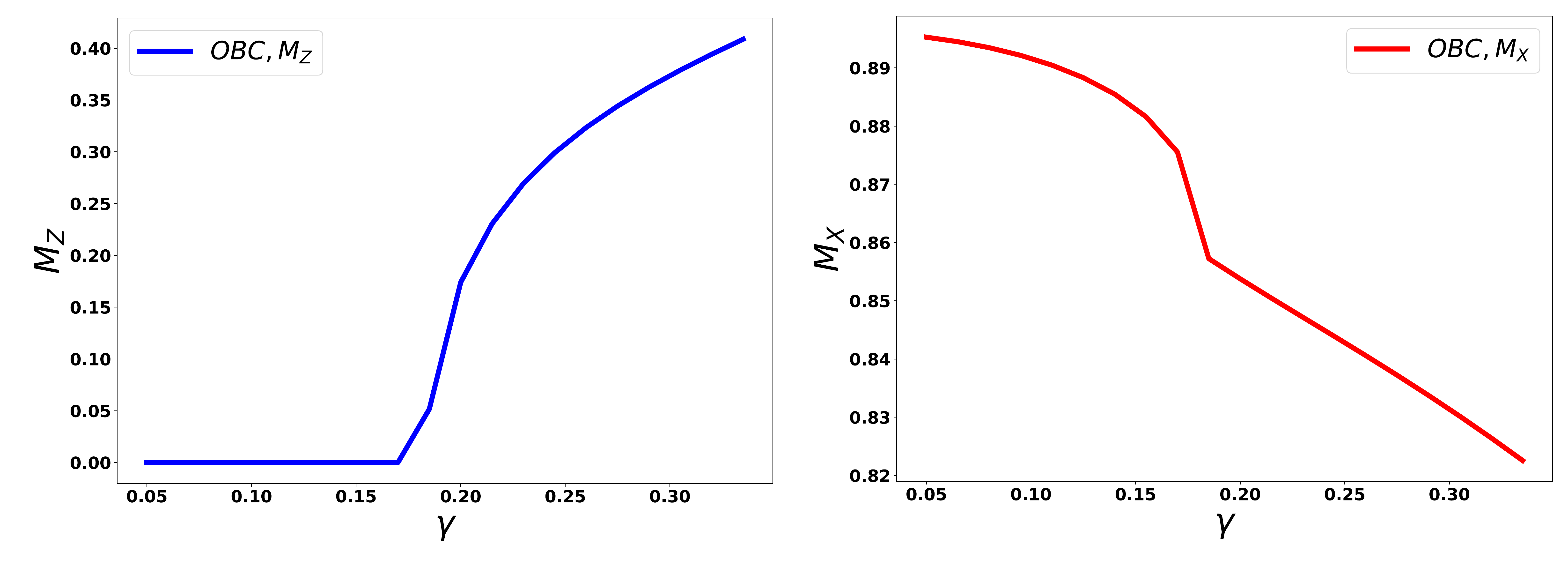}
	\caption{Order parameters $M_x$ and $M_{z}$ as defined in Eq.~\eqref{order}, which are computed by the ground state of the model from Eq.~\eqref{supptfi} under open boundary conditions (OBCs). We set $L=8$, $h_{x}=1.5$, and $J=1$. The YLES appears at $\gamma_{YL}=0.184$ where order parameters are discontinuous. Both order parameters are obtained using exact diagonalization (ED).}
	\label{fig:order}
\end{figure}

As plotted in FIG.~\ref{fig:order}, we present such two kinds of order parameters, and both of them can capture critical points $\gamma_{YL}$ which appear at the kink along $\gamma$. Since a singular value cut-off is required in the matrix product states computation discussed later, we choose the order parameter $M_x$ for identifying phase transition in this work. The divergence near critical points is not intense, and the measurement in $M_x$ helps to minimize the error in our MPS computation.

\begin{figure}[h!]
	\centering
	\includegraphics[width=0.7\linewidth]{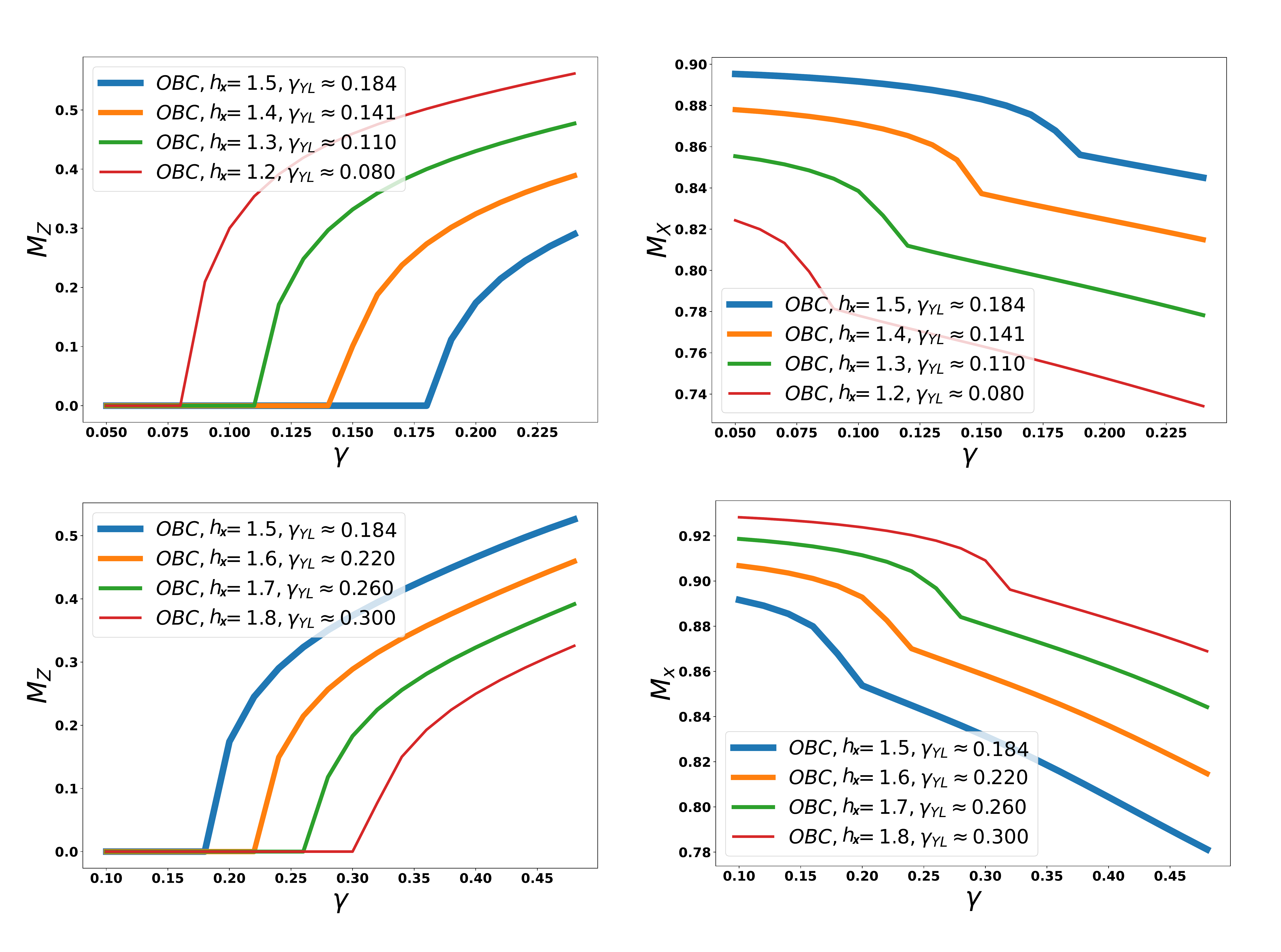}
	\caption{Order parameters $M_x$ and $M_{z}$ computed by the ground state of the model in Eq.~\eqref{supptfi} under OBCs, for different transverse fields $h_x$. We set $L=8$, and $J=1$ for all plots. Order parameters are discontinuous at critical points $\gamma_{YL}$.}
	\label{fig:mx3}
\end{figure}

Next, in FIG.~\ref{fig:mx3}, we plot order parameters computed in ground states under different transverse fields $h_x$, and the critical point $\gamma_{YL}$ appears when $h_{x}>J$. Under different $h_{x}$, we find that critical behaviors of $M_x$ and $M_{z}$ are qualitatively similar. Hence both eligible to describe critical phase transitions in our model. To avoid overly dramatic divergences near the critical points, we choose $M_x$ as our phase-transition indicator.
\subsection{Dynamical signature of phase transitions}
Due to the atom loss in our proposed setup, it is difficult to access the ground state. Below, we provide more detailed results supporting the use of a dynamical signature for detecting the phase transitions, as we have used in our main proposal.

\begin{figure}[h]
	\centering
	\includegraphics[width=0.9\linewidth]{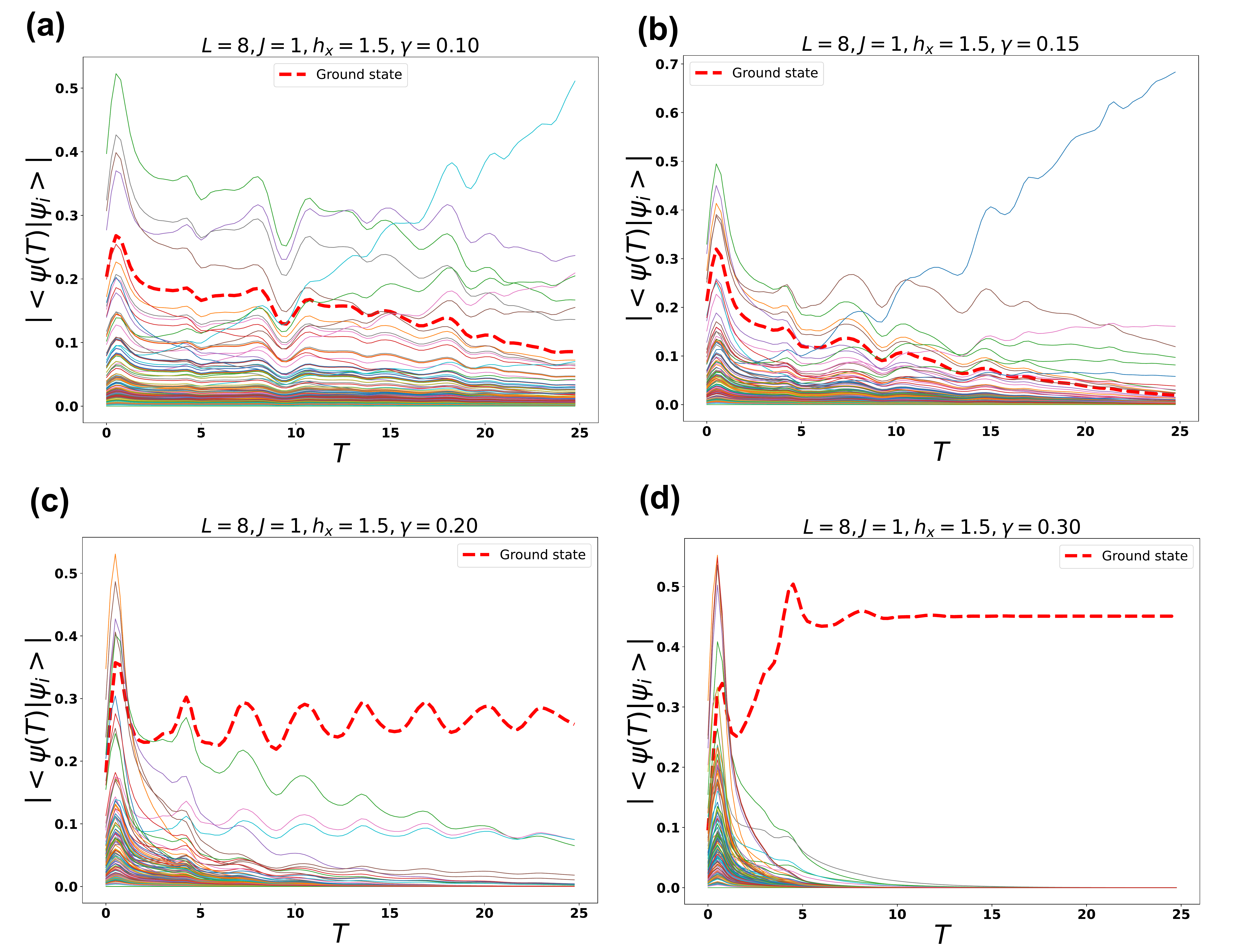}
	\caption{(a) Dynamical overlap $|\bra{\psi(t)}\ket{\psi_{i}}|$ of the evolved state $\ket{\psi(t)}\propto e^{-iT\hat H_\text{TFI}}\ket{\psi_{0}}$ with each of the eigenstates $\ket{\psi_i}$ of the Hamiltonian $\hat H_\text{TFI}$ in Eq.~\eqref{supptfi} under OBCs and $J=1$, $h_{x}=1.5$. The initial state is $\ket{\psi_{0}}=\ket{\downarrow\downarrow......}$. The red dashed curves refer to the overlap with the ground state. For (a) and (b), we set $\gamma<\gamma_{YL}$, and $\gamma>\gamma_{YL}$ for (c) and (d). When $\gamma>\gamma_{YL}$, the ground state with imaginary eigenvalue ${\rm Im}E_{g}>0$ evolves to a high overlap which is a dominated state through dynamics. The huge contrast in the dynamical behavior and overlap provides a clear signature of the phase transition shown in FIG.~\ref{fig:mx3}.}
	\label{fig:overlap3}
\end{figure}

According to the protocol introduced in the main text, we propose to measure the YLES through responses under quench dynamics $\ket{\psi(t)}\propto e^{-iT\hat H_\text{TFI}}\ket{\psi_{0}}$. To understand why such dynamics can exhibit dramatic response near critical points, we plot the dynamical overlap for every eigenstate in FIG.~\ref{fig:overlap3}. Here, the ground-state overlaps (red curves) exhibits significantly different behaviors across the phase transition. When $\gamma>\gamma_{YL}$ for the ferromagnetic phase, the ground-state overlap evolves to a high plateau value under the ferromagnetic initial state as $\ket{\psi_{0}}=\ket{\downarrow\downarrow......}$, which indicates that the ground state govern the dynamics. However, when $\gamma<\gamma_{YL}$, the ground states host real eigenvalues, and accordingly, the dynamics will be governed by other states with complex eigenvalues, as shown in FIG.~\ref{fig:overlap3} (a) and (b).

\newpage
\begin{figure}[h]
	\centering
	\includegraphics[width=0.67\linewidth]{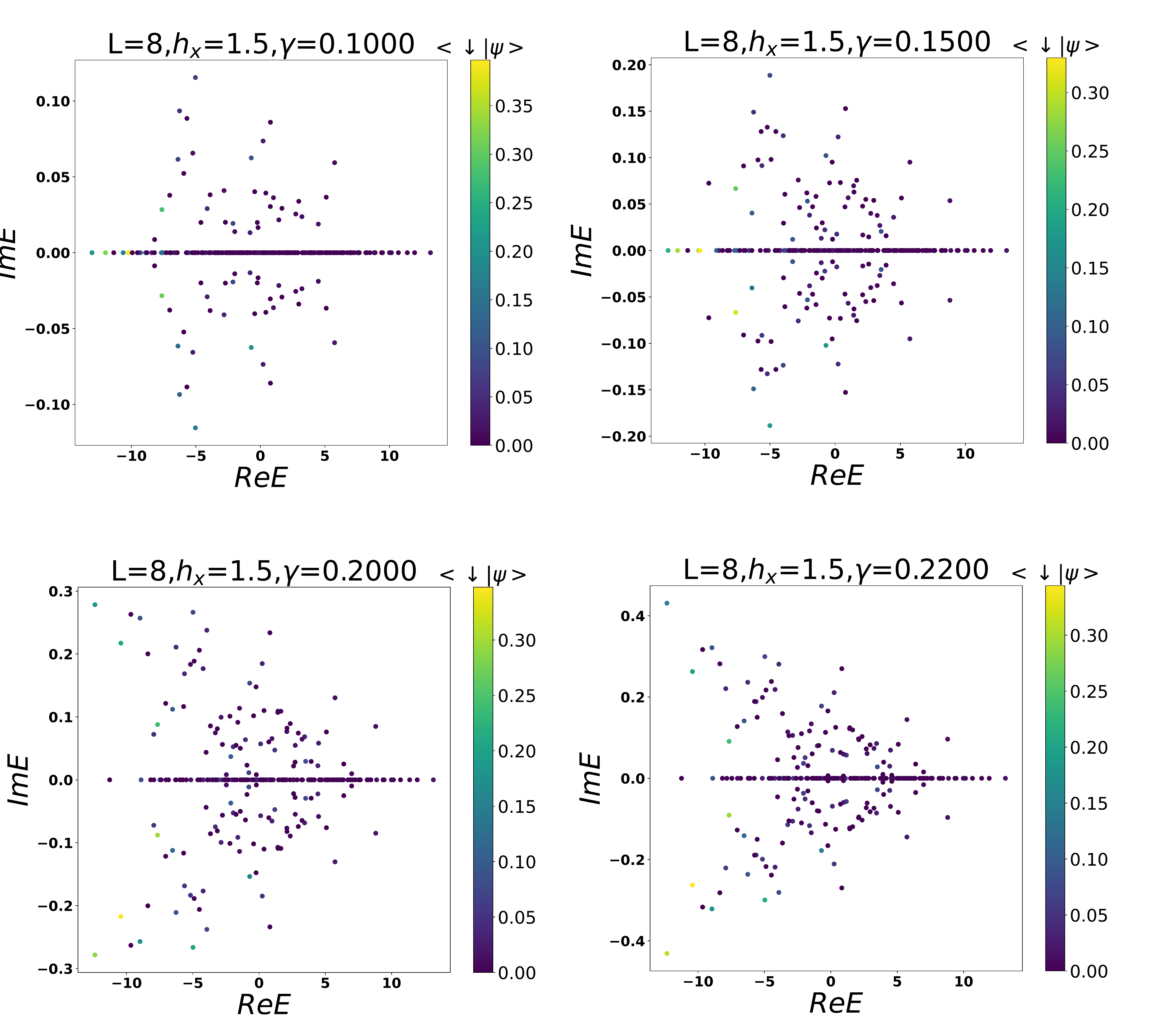}
	\caption{Spectra of the Hamiltonian in Eq.~\eqref{supptfi} under OBCs with different colors indicating the overlap between the particular eigenstate $\ket{\psi}$ and the ferromagnetic initial state $\ket{\downarrow}=\ket{\downarrow\downarrow\downarrow...}$. We set $L=8,J=1$ and $h_{x}=1.5$ which correspond to the critical point value $\gamma_{YL}=0.184$. Due to the rapid repulsion of the ground state (with smallest $\text{Re}(E)$) eigenenergies away from the real line when $\gamma>\gamma_{YL}$, the ground states evolve to those with the maximum imaginary spectra rapidly.}
	\label{fig:overlap4}
\end{figure}

To understand the above dynamical behavior under $\gamma>\gamma_{YL}$, we plot the spectra in FIG.~\ref{fig:overlap4} and check the eigenstate overlap with our prepared initial state $\ket{\downarrow\downarrow......}$  before dynamical evolution. As shown in FIG.~\ref{fig:overlap4}, the ground states host relativily high overlap with $\ket{\downarrow\downarrow......}$. In particular, one of the ground states evolves to that with the maximum imaginary eigenvalue quickly when $\gamma$ exceeds $\gamma_{YL}=0.184$ under $J=1,h_{x}=1.5, L=8$. The above facts indicate that such a ground state can be a dominating state in our designed quenching dynamics when $\gamma>\gamma_{YL}$.

\newpage
\section{III. Numerical method}

In our work, we use ED and time evolution with matrix product states (tMPS) to compute the dynamics for our model. Compared with ED, the method of tMPS is a state-of-art tool to simulate larger system sizes for generic one-dimensional quantum many-body systems~\cite{White1992,vidal2004efficient,White2004}. As presented in the main text, we simulate quenching dynamics for our model with tMPS to determine the YLES and emulate the experimental process. Here, we provide more details about our tMPS method. From our results from tMPS, we also discuss the scaling behavior of Yang-Lee edge singularities.
\subsection{Time evolution with matrix product states (tMPS)}\label{tMPS}
\begin{figure}[h!]
	\centering
	\includegraphics[width=0.75\linewidth]{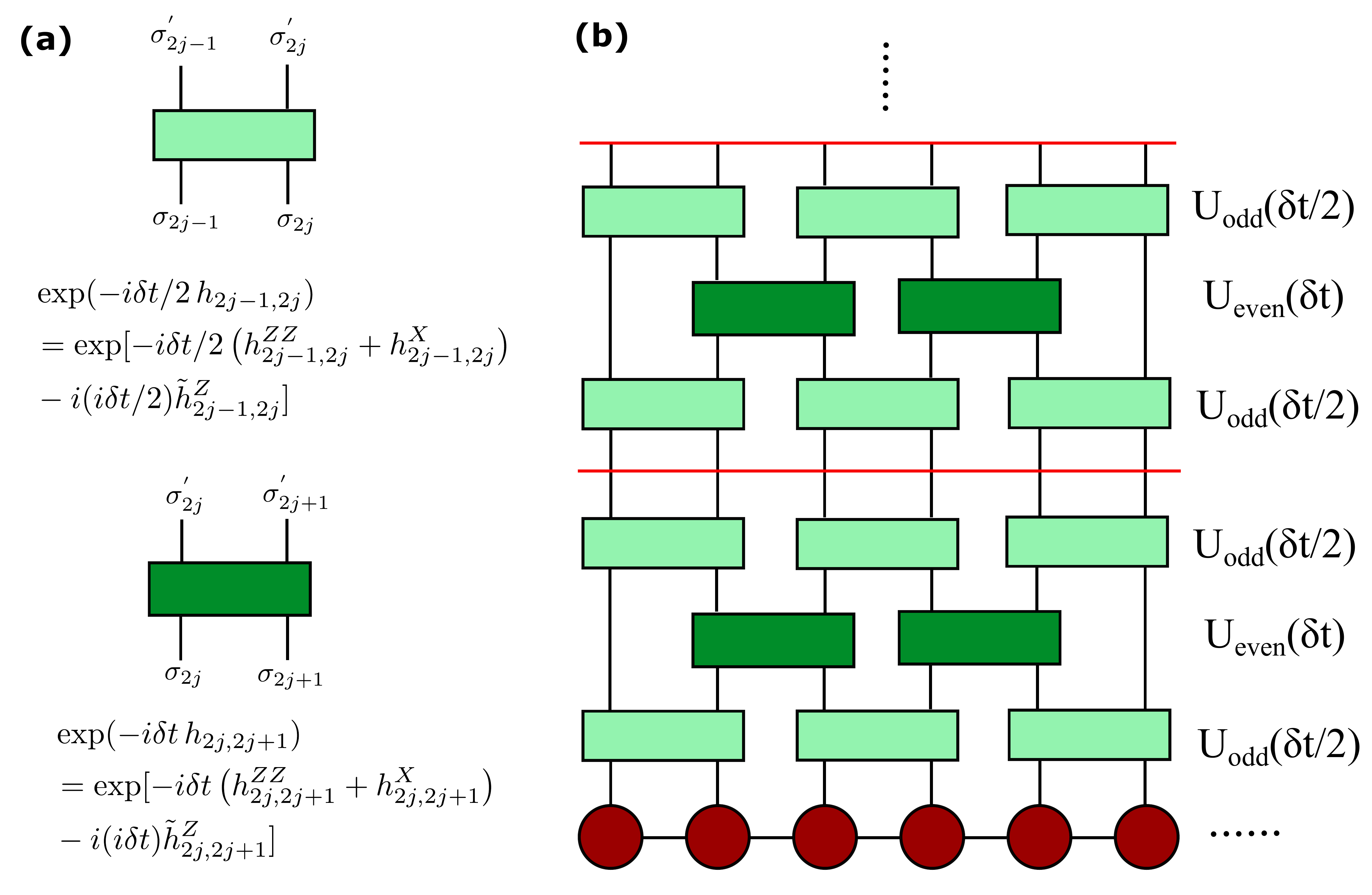}
	\caption{Time evolution with MPS (tMPS) with a second-order Suzuki-Trotter decomposition algorithm of the evolution operator: (a) Odd and even-bond time evolution operator in the form of a rank-$4$ matrix product operator (MPO). The MPO representation for the time evolution operator on the odd bond $\exp[-i\delta t/2\, h_{2j-1,2j}]$ (light green) and the even bond $\exp[-i\delta t/2\, h_{2j,2j+1}]$ (dark green) has four indices, of which each has a dimension of two. The indices without prime are contracted with the MPS in the time evolution. (b) Two adjacent steps of tMPS. The initial state is prepared in the right-canonical form (red). The total time evolution operator for each time step $U(\delta t)$ is decomposed into two odd-bond parts $U_{\text{odd}}(\delta t/2)$ (light green) and one even-bond part $U_{\text{even}}(\delta t)$ (dark green) by a symmetrical decomposition. At the end of each time step $\delta t$, a normalization of the state (solid red line) is performed.}
	\label{fig:mps}
\end{figure}

Here, we treat the total Hamiltonian evolution as a hybrid real-time and imaginary-time evolution and apply a second-order Suzuki-Trotter decomposition~\cite{Suzuki1990} to each trotter step $U(\delta t)$, after which this time evolution operator $U(\delta t)$ can be written as [see Fig.~\ref{fig:mps}(b)]
\begin{align}
	\label{su}
	&U(\delta t)= U_{\text{odd}}(\delta t/2)U_{\text{even}}(\delta t)U_{\text{odd}}(\delta t/2) + \mathcal{O}(\delta t^3),
\end{align}
where
\begin{equation}
	\begin{aligned}
		&U_{\nu}(\delta t)=\exp\left[-i\delta t \left(H_{ZZ}^{\nu}+H_X^{\nu}\right) - i\left(i\delta t\right)\tilde{H}_Z^{\nu}\right] (\nu\text{ is either odd or even}).
	\end{aligned}
\end{equation}

The terms in the above equation are explicitly:
\begin{equation}
	\begin{aligned}
		&H_{ZZ}^{\text{odd}}=\sum_{j\,\,\text{odd}}h_{j,j+1}^{ZZ}=\sum_{j=1}^{N/2}-J \sigma_{2j-1}^z\otimes\sigma_{2j}^z,\\
		&H_{ZZ}^{\text{even}}=\sum_{j\,\,\text{even}}h_{j,j+1}^{ZZ}=\sum_{j=1}^{N/2-1}-J \sigma_{2j}^z\otimes\sigma_{2j+1}^z, \\
		&H_X^{\text{odd}}=\sum_{j\,\,\text{odd}}h_{j,j+1}^{X}=(-h_{x})\sigma_1^x \otimes \mathds{1}_2 + \mathds{1}_1 \otimes (-h_{x}/2)\sigma_2^x +\sum_{j=1}^{N/2-1}(-h_{x}/2)\sigma_{2j-1}^x \otimes \mathds{1}_{2j} + \mathds{1}_{2j-1}\otimes(-h_{x}/2)\sigma_{2j}^x\\
		&+(-h/2)\sigma_{N-1}^x \otimes \mathds{1}_N + \mathds{1}_{N-1}\otimes(-h)\sigma_N^x,\\
		&H_X^{\text{even}}=\sum_{j\,\,\text{even}}h_{j,j+1}^{X}= \sum_{j=1}^{N/2-1}(-h_{x}/2)\sigma_{2j}^x\otimes\mathds{1}_{2j+1}+\mathds{1}_{2j}\otimes(-h_{x}/2)\sigma_{2j+1}^x\\
		&\tilde{H}_Z^{\text{odd}} =\sum_{j\,\,\text{odd}}\tilde{h}_{j,j+1}^{Z}= \gamma\sigma_1^z \otimes \mathds{1}_2 + \mathds{1}_1 \otimes (\gamma/2)\sigma_2^z +\sum_{j=1}^{N/2-1}(\gamma/2)\sigma_{2j-1}^z \otimes \mathds{1}_{2j} + \mathds{1}_{2j-1}\otimes(\gamma/2)\sigma_{2j}^z,\\
		&+(\gamma/2)\sigma_{N-1}^z \otimes \mathds{1}_N + \mathds{1}_{N-1}\otimes \gamma\sigma_N^z,\\ \\
		&\tilde{H}_Z^{\text{even}}=\sum_{j\,\,\text{even}}\tilde{h}_{j,j+1}^{Z} = \sum_{j=1}^{N/2-1}(\gamma/2)\sigma_{2j}^z\otimes\mathds{1}_{2j+1}+\mathds{1}_{2j}\otimes(\gamma/2)\sigma_{2j+1}^z,
	\end{aligned}
\end{equation}
and therefore $U_{\text{odd}}(\delta t/2)$ or $U_{\text{even}}(\delta t)$ can be re-written as a product of local two-body terms [see Fig.~\ref{fig:mps}(a)]:
\begin{align}
	&U_{\text{odd}}(\delta t/2)=\exp\left[-i\delta t/2 \left(H_{ZZ}^{\text{odd}}+H_X^{\text{odd}}\right) - i\left(i\delta t/2\right)\tilde{H}_Z^{\text{odd}}\right] \\ \nonumber &=\exp\left[-i\delta t/2\left(\sum_{j\,\, \text{odd}}h_{j,j+1}^{ZZ}+\sum_{j\,\,\text{odd}}h_{j,j+1}^X \right)-i(i\delta t/2)\sum_{j\,\,\text{odd}}\tilde{h}_j^Z\right] \\ \nonumber &=\prod_{j\,\,\text{odd}} \exp\left[-i\delta t/2 \left(h_{j,j+1}^{ZZ}+h_{j,j+1}^X\right)-i\left(i\delta t/2\right)\tilde{h}_j^Z\right] \\ \nonumber
	&=\prod_{j\,\,\text{odd}}\exp\left[-i\delta t/2\, h_{j,j+1}\right],\\ \nonumber
	&U_{\text{even}}(\delta t)=\exp\left[-i\delta t \left(H_{ZZ}^{\text{even}}+H_X^{\text{even}}\right) - i\left(i\delta t\right)\tilde{H}_Z^{\text{even}}\right] \\ \nonumber &=\exp\left[-i\delta t\left(\sum_{j\,\, \text{even}}h_{j,j+1}^{ZZ}+\sum_{j\,\,\text{even}}h_{j,j+1}^X \right)-i(i\delta t)\sum_{j\,\,\text{even}}\tilde{h}_j^Z\right] \\ \nonumber &=\prod_{j\,\,\text{even}} \exp\left[-i\delta t \left(h_{j,j+1}^{ZZ}+h_{j,j+1}^X\right)-i\left(i\delta t\right)\tilde{h}_j^Z\right] \\ \nonumber
	&=\prod_{j\,\,\text{even}}\exp\left[-i\delta t\,h_{j,j+1}\right].
\end{align}
For our setup, the system size $L$ is even, and we have explicitly written both odd-bond and even-bond Hamiltonians as two-body tensor products of local spin operators. $\mathds{1}_j$ is the local identity operator in the spin-$1/2$ basis. We have also introduced $\tilde{H}_{Z}^{\text{odd}}$ and $\tilde{H}_{Z}^{\text{even}}$ so that the imaginary $i$ from the imaginary field of the original $H_Z$ has been absorbed into the imaginary time evolution step $i\delta t$. Considering a long time $T$ which is discretized into $T/\delta t$ steps, if applying the $U(\delta t)$, the error accumulates to $T/\delta t \mathcal{O}(\delta t^3)\approx \mathcal{O}(\delta t^2)$ for the total time interval $T$, and therefore it renders a second-order time evolution. Since the implementation of tMPS algorithm with PBC at this stage is computationally expensive, we only consider the OBCs for the tMPS results in this work. Based on our MPS method, we realize the simulation of long-time dynamics ($T=30$), as shown in FIG.~\ref{fig:mt}, which are difficulty for other ancilla-based methods ~\cite{lin2021real,chen2022high}. According to the results in FIG.~3 (a) from main text, our MPS simulation can give the precise postions of critical points.

\subsection{Additional numerical results}
\begin{figure}[h!]
	\centering
	\includegraphics[width=0.7\linewidth]{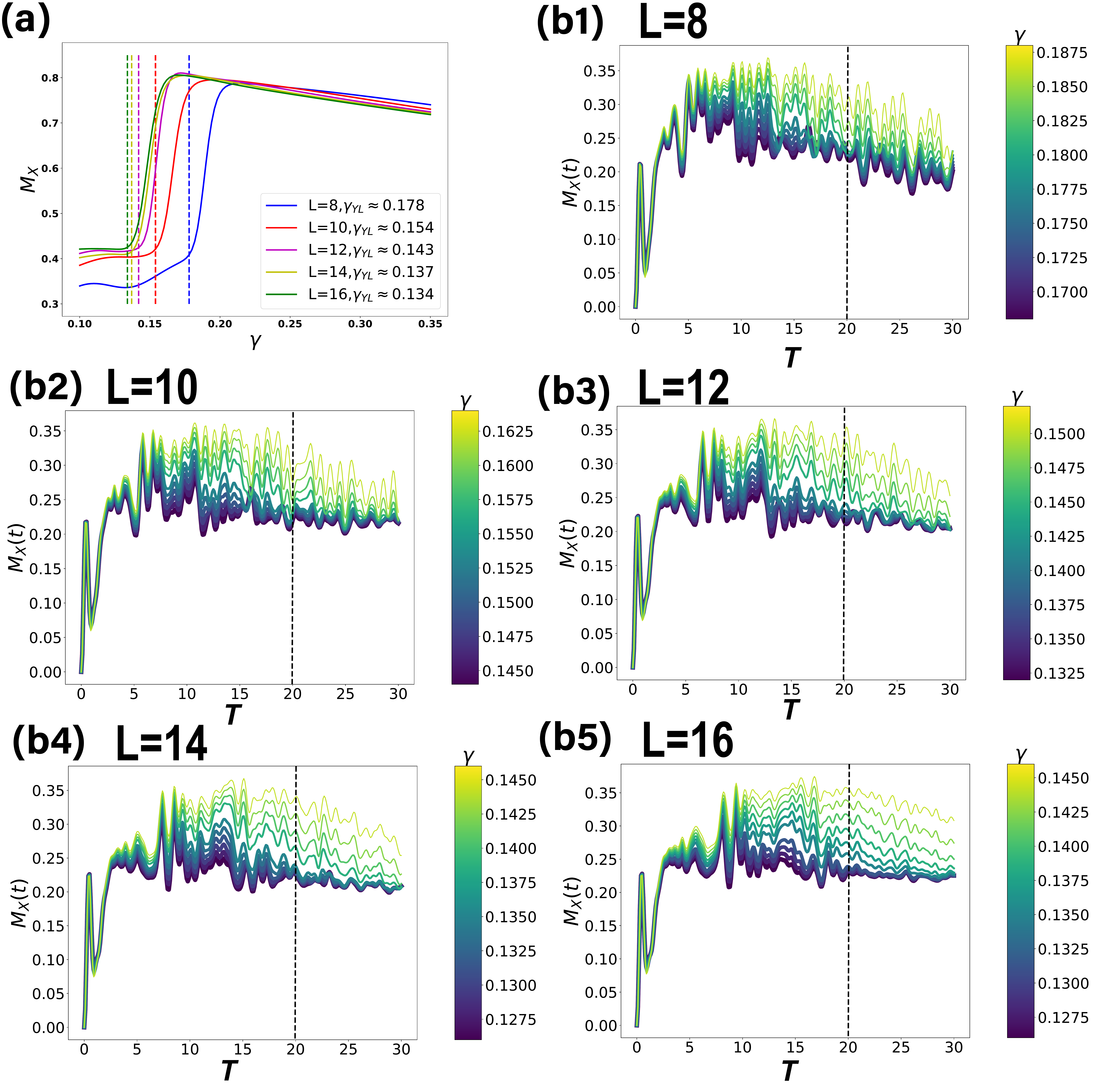}
	\caption{(a) Dynamical order parameter $M_x(T)$ measured at time $T=20$ under $J=1$ (black dashed lines marked in (b1)-(b5)) with different $\gamma$. We set parameters as $J=1$, $h_{x}=1.5$, and  $L=8,10,12,14,16$, and prepare the initial state as $\ket{\downarrow\downarrow\downarrow\downarrow\downarrow\downarrow\downarrow\downarrow}$ under OBCs. Order parameters exhibit intense growth near critical points $\gamma_{YL}$ marked by dashed lines. (b1)-(b5) Dynamical order parameter $M_x(T)$ with initial state $\ket{\downarrow\downarrow\downarrow\downarrow\downarrow\downarrow\downarrow\downarrow}$ under different sizes $L$ and $\gamma$. We set $JT=20$ as the measuring time (black dashed lines) for plot (a) and the main text such that we can distinctively see the curves spreading out for $\gamma>\gamma_{YL}$ but not $\gamma<\gamma_{YL}$.
	}
	\label{fig:mt}
\end{figure}

As described in the main text, we propose to measure the dynamical order parameter as $M_x(T)=|\bra{\psi(T)}\frac{\sum_{j}\sigma^{x}_{j}}{L}\ket{\psi(T)}|$ at a fixed time $T$ and fixed $h$ and $J$, such that our suggested order parameter $M_x(T)$ diverges when $\gamma$ sweeps past the critical point $\gamma_{YL}$. For this, we need to choose the duration of evolution $T$ for $M_x(T)$ properly. Here, we suggest measuring the order parameter at time $JT=20$ with $J=1$, and supporting insights are shown in FIG.~\ref{fig:mt} (b1) -(b5), where we present the full picture of the dynamics under different sizes $L$. A duration $T=20$ ensures that the curves spread out for $\gamma>\gamma_{YL}$ but not $\gamma<\gamma_{YL}$ for all sizes $L$. As such, measuring at $T=20$ can reveal the mentioned divergence in the dynamical order parameter at $\gamma_{YL}$.

\newpage
\subsection{Additional results for experimental implementations}
\begin{figure}[h]
	\centering
	\includegraphics[width=0.7\linewidth]{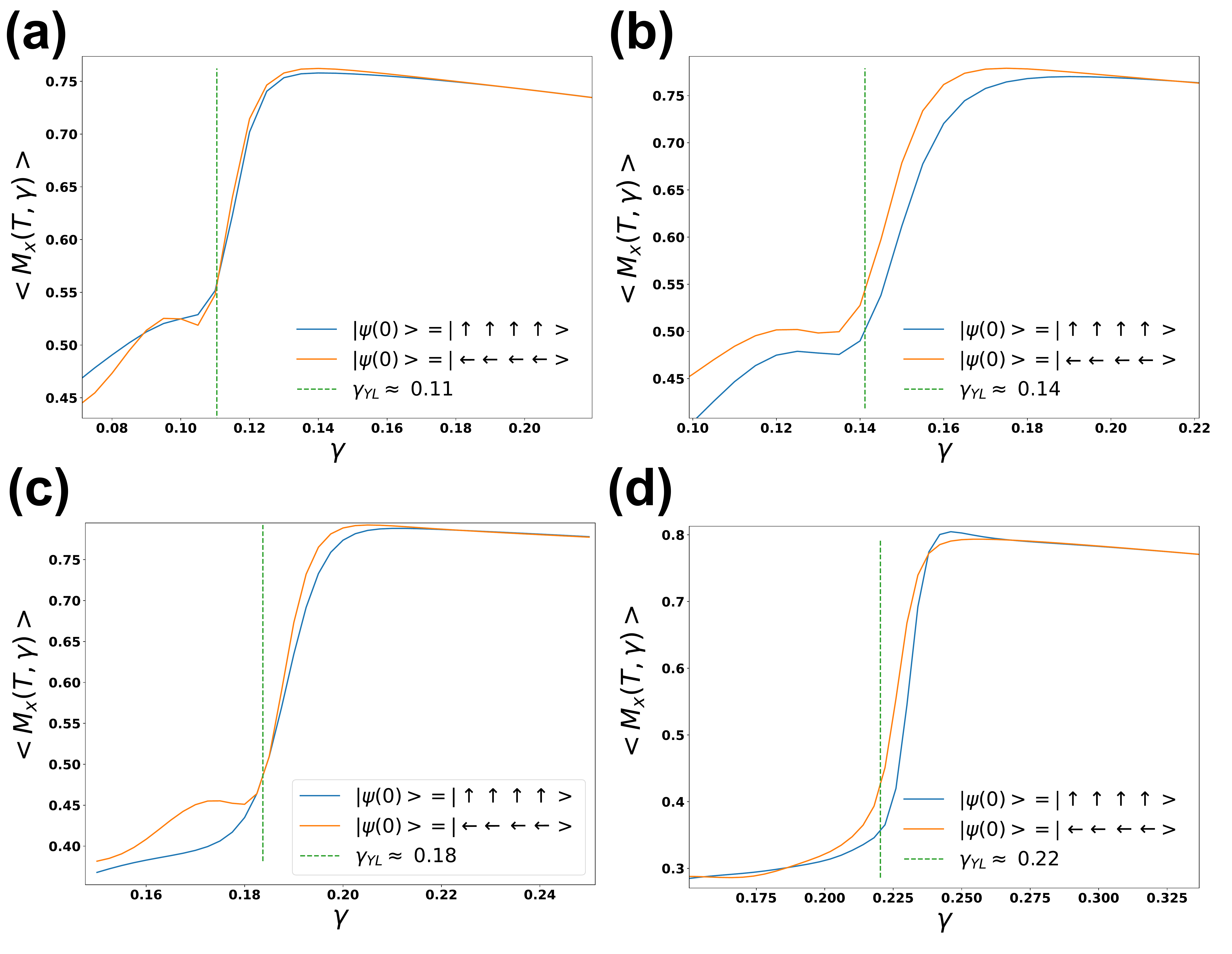}
	\caption{Dynamical order parameter $M_x(T,\gamma)$ measured at time $T=20$ along $\gamma$. The initial state are: $\ket{\downarrow\downarrow\downarrow\downarrow\downarrow\downarrow\downarrow\downarrow}$ for the blue curves and $\ket{\leftarrow\leftarrow\leftarrow\leftarrow\leftarrow\leftarrow\leftarrow\leftarrow}_{y}$ for the yellow curves. The green dashed lines mark the critical points $\gamma_{YL}$. Other parameters are $J=1$, $L=8$, and $h_{x}=1.3, 1.4, 1.5, 1.6$.}
	\label{fig:mxmz}
\end{figure}
In our proposed experiment scheme for observing the Yang-Lee phase transitions, we suggested the preparation of $\ket{\leftarrow\leftarrow\leftarrow...\leftarrow}_{y}$ as the initial state. In FIG.\ref{fig:mxmz}, we compare the results of $M_x(T,\gamma)$ through dynamics $e^{-iT \hat{H}_{\rm TFI}}\ket{\leftarrow\leftarrow\leftarrow...\leftarrow}_{y}$ and $e^{-iT \hat{H}_{\rm TFI}}\ket{\downarrow\downarrow\downarrow...\downarrow}$. Here there is no significant difference between the results under these two iniital states. In particular, these two initial states give similar positions of the critical points $\gamma_{YL}$.

  \end{document}